\documentclass[acmlarge,screen]{acmart}

\usepackage{subfiles}
\usepackage{xcolor}
\usepackage{multirow}
\usepackage[caption=false]{subfig}
\usepackage[export]{adjustbox}
\usepackage{url}
\usepackage{soul}
\usepackage[mathscr]{eucal}
\usepackage{commath}
\usepackage{amsmath}
\usepackage{xspace}
\usepackage[font=small]{caption}
\usepackage{mathtools}
\usepackage{graphicx}
\usepackage{parcolumns}
\usepackage{listings}
\usepackage{array}
\newcolumntype{P}[1]{>{\centering\arraybackslash}p{#1}}

\AtBeginDocument{%
  \providecommand\BibTeX{{%
    \normalfont B\kern-0.5em{\scshape i\kern-0.25em b}\kern-0.8em\TeX}}}

\setcopyright{acmcopyright}
\acmJournal{IMWUT}
\acmYear{2020} \acmVolume{4} \acmNumber{3} \acmArticle{82} \acmMonth{9} \acmPrice{15.00}\acmDOI{10.1145/3411808}

\newcommand{\Sys}{Zygarde\xspace}

\newcommand\hnote[1]{\textcolor{black}{#1}}
\newcommand{\parlabel}[1]{\vspace{0.5em}\noindent\textbf{#1}}

\begin{document}

\title{\Sys: Time-Sensitive On-Device Deep Inference and Adaptation on Intermittently-Powered Systems}

\author{Bashima Islam}
\affiliation{%
  \institution{University of North Carolina at Chapel Hill}
  \streetaddress{201 S. Columbia St}
  \city{Chapel Hill}
  \country{USA}}
\email{bashima@cs.unc.edu}

\author{Shahriar Nirjon}
\affiliation{%
  \institution{University of North Carolina at Chapel Hill}
  \streetaddress{201 S. Columbia St}
  \city{Chapel Hill}
  \country{USA}}
\email{nirjon@cs.unc.edu}

\renewcommand{\shortauthors}{Islam and Nirjon}

\begin{abstract}
We propose \emph{\Sys} --- which is an energy- and accuracy-aware soft real-time task scheduling framework for batteryless systems that flexibly execute deep learning tasks\footnote{The DNN, by definition, refers to neural networks having more than one hidden layers~\cite{hanin2017universal,lu2017expressive,hornik1991approximation,cybenko1989approximation}. Thus, a wide variety of networks qualify as a DNN in the existing literature. DNNs considered in this paper have up to $10^5$ neurons and weights combined. They fit into 256KB memory of an MCU; have convolutional, ReLU, pooling, and fully-connected structures as regular DNNs; and perform on-device inference~\cite{gobieski2019intelligence, lee2019neurozero, lee2019learning}} that are suitable for running on microcontrollers. The sporadic nature of harvested energy, resource constraints of the embedded platform, and the computational demand of deep neural networks (DNNs) pose a unique and challenging real-time scheduling problem for which no solutions have been proposed in the literature. We empirically study the problem and model the energy harvesting pattern as well as the trade-off between the accuracy and execution of a DNN. We develop an \textit{imprecise computing}-based scheduling algorithm that improves the timeliness of DNN tasks on intermittently powered systems. We evaluate \Sys using four standard datasets as well as by deploying it in six real-life applications involving audio and camera sensor systems. Results show that \Sys decreases the execution time by 
up to 26\% and schedules 
9\% -- 34\% more tasks with up to 21\%
higher inference accuracy, compared to traditional schedulers such as the earliest deadline first (EDF).
\end{abstract}



\maketitle

\section{Introduction}
\label{sec:introduction}


Batteryless IoT devices are powered by harvesting energy from ambient sources, such as solar, thermal, kinetic, and radio-frequency signals (RF). These devices, in principle, last forever---as long as the energy harvesting conditions are in favor. They have applications in long-term sensing and inference scenarios, such as wildlife monitoring, remote surveillance, environment and infrastructure monitoring, and health monitoring using wearables and implantables. While every IoT application has an expected response time, many of them \emph{require} timely feedback. For instance, in acoustic sensing systems, such as hearables and voice assistants~\cite{hoy2018alexa, islam2017soundsifter}, car detectors~\cite{de2018paws}, and machine monitors~\cite{srinivasan2017preventive}, events need to be detected and reported on time in order to ensure prompt responses, safety, and timely maintenance. Though a batteryless system is desirable, the unpredictability of the harvested energy, combined with the complexity of on-device event recognition tasks, complicates timely execution of machine learning-based event detection tasks on such systems. 

Prior works on time-aware batteryless computing systems can be broadly categorized into two types. The first category focuses on maintaining a reliable system clock when the power is out~\cite{de2020reliable, rahmati2012tardis, hester2016persistent}. The second category includes runtime systems that consider the temporal aspect of data across power failures~\cite{hester2017timely, buettner2011dewdrop, zhu2012deos, yildirim2018ink} by discarding data after a predefined interval or considering energy to increase the chances of task completion. Recent works~\cite{gobieskiintermittent, gobieski2019intelligence, lee2019neurozero} on batteryless systems have proposed frameworks and runtime to execute \emph{non-real-time} DNN$^1$ tasks on intermittently-powered systems. The real-time community have proposed techniques~\cite{bateni2018apnet, zhou2018s, yang2019re, lee2020subflow, islam2020scheduling} for deadline-aware execution of DNNs, primarily on GPU and server-grade machines. Despite these commendable efforts, there is a gap in the existing literature that none has considered all three dimensions, i.e., intermittence of harvested energy, variable utility of DNN inference, and real-time schedulability; and merely combining existing solutions does not entirely solve the problem. 



\begin{figure}[!t]
    \centering
    \includegraphics[width=0.9\textwidth]{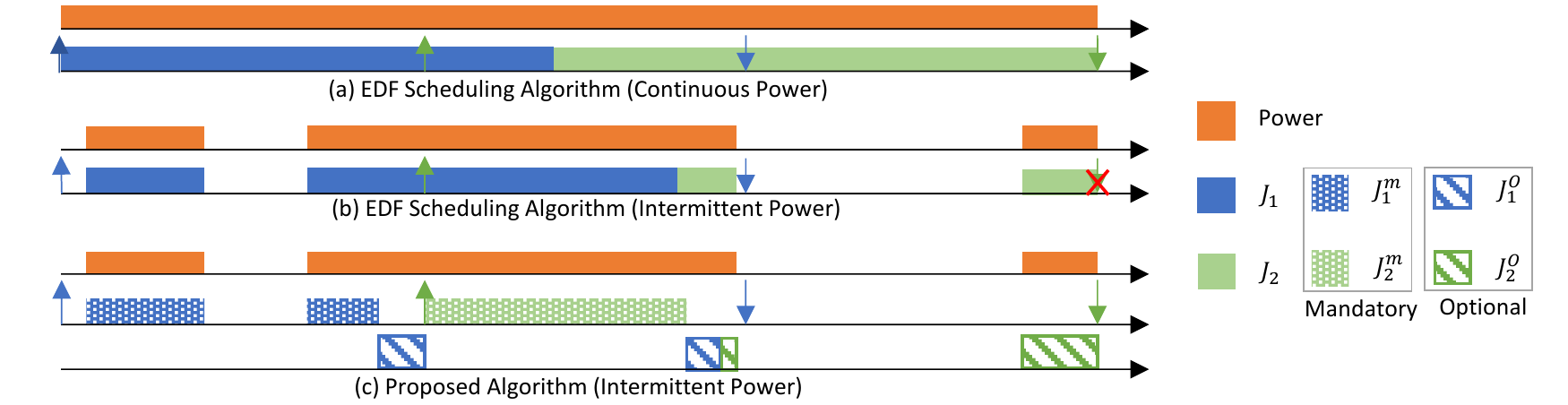}
    \vspace{-10pt}
    \caption{(a) With constant power both tasks meet their deadline. (b) With intermittent power, job $J_2$ misses its deadline. (c) \hnote{When tasks are imprecise, the mandatory parts of both jobs complete on time, and some optional part of $J_1$ gets done as well.} }
    \label{fig:intermittent_deadmiss}
    \vspace{-10pt}
\end{figure}


We illustrate this using an example in Figure~\ref{fig:intermittent_deadmiss}. We consider two jobs, $J_1$ and $J_2$, released at time 0 and 20, respectively. Their relative deadline is 34, and the execution time is 28. The intermittency of energy generally has no consistent pattern in the duration of or in the gaps between ON/OFF phases. Figure~\ref{fig:intermittent_deadmiss}(a) shows that when the power is uninterrupted, both jobs meet their deadlines under the earliest deadline first (EDF) scheduling. When power is intermittent (Figure~\ref{fig:intermittent_deadmiss}(b)), task $J_2$ misses its deadline. Figure~\ref{fig:intermittent_deadmiss}(c) illustrates our proposed approach which partitions a job into mandatory and optional portions and ensures that the mandatory portion (which is required to achieve a desirable inference accuracy) of each job finishes on time. And if there is extra time, our proposed approach schedules some optional jobs that may increase the accuracy further.

In this paper, we present \Sys --- which is the first system that enables deadline-aware imprecise execution of DNNs on an intermittently-powered system. The design of \Sys is motivated by two observations. First, energy generated by a harvester is \emph{bursty}, i.e., the state of energy generation is maintained over a short period. This property enables us to obtain a probabilistic model of the energy harvesting pattern -- which can be characterized by a single parameter for a given harvester used in a particular application scenario. Second, since deeper layers of a DNN extract more detailed and fine-grained features of the input data, for a given accuracy, the required amount of DNN computation depends on the quality of data itself. \Sys exploits these observations and
proposes an \textit{imprecise computing}-based~\cite{shih1992line, liu1991algorithms} online scheduling algorithm, which considers both the intermittence of energy and the accuracy-execution trade-off of a DNN. 

The main contribution of this paper is a deadline-aware runtime framework for DNNs executing on intermittently-powered systems, for which, runtime adaptation of a DNN is necessary, on top of compile-time compression~\cite{gobieski2019intelligence, han2015deep,nabhan1994toward, yao2017deepiot, yao2018rdeepsense, yao2018apdeepsense, yao2018fastdeepiot}. Compile-time compression alone is not sufficient when the remaining deadline is inadequate for a full execution of the DNN, but long enough to compute the inference result from the partial execution of the DNN. This runtime adaptation process requires (1) modeling the energy harvesting pattern using a single factor($\eta$), which helps determine the system how much computation is possible in the near future, and (2) specialized offline training of the DNNs to minimize the loss of accuracy due to partial execution.

\Sys complements prior work on batteryless systems such as time-keeping~\cite{rahmati2012tardis, hester2016persistent} and execution of non-real-time tasks~\cite{gobieskiintermittent, gobieski2019intelligence, ransford2012mementos, mirhoseini2013automated,balsamo2015hibernus, balsamo2016hibernus++,maeng2017alpaca, colin2016chain, maeng2018adaptive}. In contrast to all previous works, \Sys's contribution is at the framework, modeling, and algorithmic level, while its implementation relies upon existing open-source frameworks and APIs~\cite{gobieski2019intelligence, maeng2017alpaca} that handle the lower-level aspects of an intermittent system. 

We implement \Sys on a TI MSP430FR5994 microcontroller and evaluate its performance using four standard datasets (MNIST~\cite{lecun1998mnist}, ESC-10~\cite{piczak2015environmental}, CIFAR-100~\cite{krizhevsky2009learning}, and Visual Wake Works~\cite{chowdhery2019visual}) as well as in six real-world acoustic event detection and visual sensing experiments. \Sys achieves 5\%--26\% reduction in execution time using early termination. By using the layer-aware loss for early termination, it also increases the inference accuracy by upto 21\% than the state-of-the-art solutions that use cross-entropy loss ~\cite{zhang2018generalized} and contrastive loss~\cite{islam2019soundsemantics}. Moreover, it schedules 9\%--34\% more jobs with upto 30\% higher inference accuracy than the earliest deadline first (EDF) scheduling algorithm. Furthermore, it gains up to 28\% higher inference accuracy than the imprecise variant of EDF.
\section{\Sys~System Design}
\label{sec:overview}

In this section, we describe the system architecture of \Sys and provide an example to illustrate its execution.   

\subsection{System Architecture}

\begin{figure}[!thb]
    \centering
    \includegraphics[width=.8\textwidth]{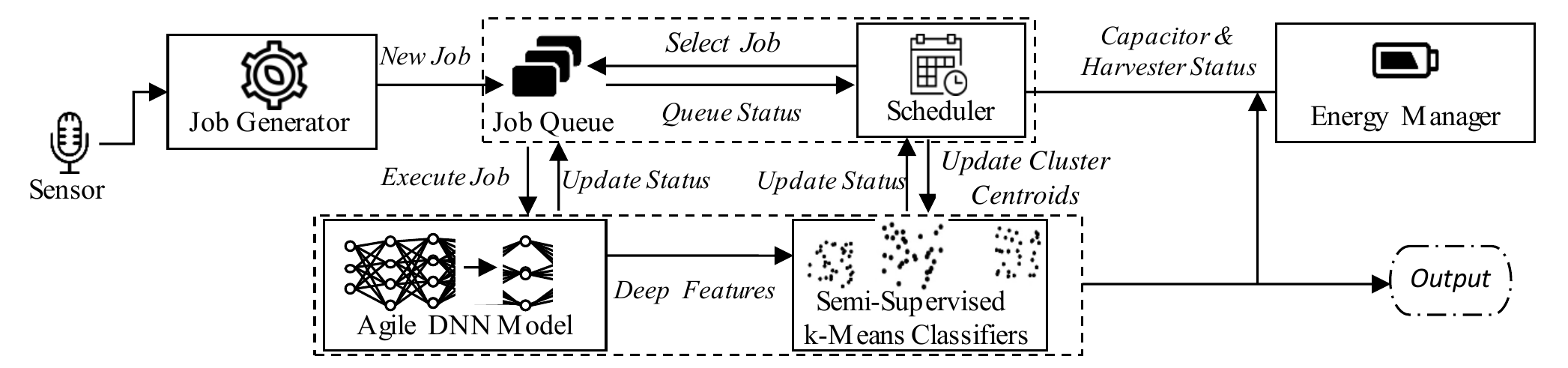}
    \vspace{-5pt}
    \caption{\Sys System Architecture.}
    \vspace{-15pt}
    \label{fig:sys}
\end{figure}

\Sys is a framework and a runtime system that enables time-aware execution of a DNN on intermittently-powered systems. It consists of five major components: a job generator, an energy manager, an agile DNN model, semi-supervised $k$-means classifiers, and a scheduler. These components are shown in Figure~\ref{fig:sys}.  

\parlabel{Job Generator.} \Sys reads data from sensors (e.g., a microphone) and writes them to the non-volatile memory (FRAM) of the microcontroller using direct memory access (DMA). It considers the processing of the sensor data stream for each classification task (e.g., user identification and event detection) as separate \emph{tasks}. The end-to-end processing of one data sample of a sensor stream (i.e., from feature extraction to classification and model adaptation) is called a \emph{job}. For example, an audio sensing system that performs both speaker recognition and hotword detection has two tasks, and the processing of each audio frame to recognize the speaker and to detect the hotword are two separate jobs. Thus, if this system (having 2 tasks) generates $k$ audio frames/second, then after 3 seconds, there will be a total of $6k$ jobs. The \emph{Job Generator} creates and enqueues jobs into a queue. A job leaves the queue when it gets scheduled for execution or its deadline has passed.

\parlabel{Energy Manager.} The \emph{Energy Manager} monitors the state of the energy storage (i.e., the supercapacitor) and the energy harvesting rate. The scheduler uses this information for scheduling decisions (described in Section~\ref{sec:scheduler}). The energy manager implements an open-source intermittent computing runtime~\cite{gobieski2019intelligence, maeng2017alpaca} to manage the execution of jobs across power failures. At the implementation level, each job of \Sys consists of multiple small atomic fragments that maintain a strict precedence order. These fragments execute atomically and the runtime ensures that repeated attempts to execute a fragment is idempotent.


\parlabel{Agile DNN Model.}
The \emph{Agile DNN Model} is a pre-trained deep neural network that converts data samples into feature representations~\cite{bengio2013representation}. The network is trained offline, on a high-end server, using labeled training data. We use rank-decomposition~\cite{bhattacharya2016sparsification} and separation~\cite{han2015deep} to compress and fit the network into the limited memory of a microcontroller. Based on the quality of the input data sample, \Sys may terminate the execution of the DNN early at runtime, and hence, we call it an \emph{agile} DNN Model. We note that an agile DNN is a special type of anytime DNN~\cite{teerapittayanon2016branchynet,yu2019universally, karayev2012timely, karayev2014anytime, hu2019learning, kim2018nestednet} --- where the depth of a neural network is dynamically adjusted during inference. However, unlike anytime networks, where the output of a hidden layer is fed to a secondary shallow neural network for classification, agile DNN employs a cluster-based classifier to replace expensive matrix multiplication operations with 4X less costly additions/subtractions~\cite{addgcc, iqMathLib}. In our implementation of agile DNN, this design saves 27,750 execution cycles per inference, when compared to anytime neural networks.   



\parlabel{Semi-Supervised $k$-Means Classifiers.} The feature representation obtained from the execution of the Agile DNN is classified by a semi-supervised $k$-means clustering algorithm~\cite{arthur2006k}. The clustering algorithm uses the L1 distance between two feature vectors~\cite{black1998dictionary}. For layers (e.g., the convolution layers) that produce two or more dimensional features, they are flattened or vectorized prior to computing the L1 norm. Since the execution of an Agile DNN may terminate at any layer depending on the input data, \Sys maintains a separate $k$-means classifier for each layer of the Agile DNN. These $k$-means classifiers are trained offline on a server machine. However, to enable online learning, these classifiers are updated at runtime using a model adaptation process described in Section~\ref{sec:cluster_construction}. The motivation behind the $k$-mean based approach is to reduce the execution time and energy consumption by avoiding multiplications which is over 4$\times$ more expensive than additions and subtractions.



\parlabel{Scheduler.} \Sys implements an online, dynamic priority, real-time scheduler that considers not only the timing aspects but also the expected inference accuracy of a job and the energy harvesting status of the system. It dynamically partitions an executing job into mandatory and optional portions based on the early termination of an agile DNN and prioritizes the execution of its mandatory portion to ensure both timeliness and accuracy under the constraints of intermittently available energy. Note that, the beginning portion of all jobs are mandatory and whether the next unit is mandatory or optional is determined during the execution of the current unit.
An illustration of the scheduler is described next. 



\vspace{-5pt}
\begin{figure}[!htb]
\begin{minipage}{.5\textwidth}
\captionsetup{type=table} 
\caption{Description of the workload.}
\vspace{-10pt}
\begin{footnotesize}
\begin{tabular}{|c|c|c|c|c|c|}
\hline
\bf{Job} & \bf{Total Layers} & \bf{Mandatory} & \bf{Optional} & \bf{Release Time} & \bf{Deadline} \\ \hline
$J_{1,1}$ & 4 & 1 & 3 & $t_1$ & $t_7$ \\ 
$J_{1,2}$ & 4 & 2 & 2 & $t_3$ & $t_9$ \\ \hline
\end{tabular}
\end{footnotesize}
\label{tab:load_des}
\end{minipage}

\begin{minipage}{0.4\textwidth}
\centering
\vspace{.5em}
\includegraphics[width=\textwidth]{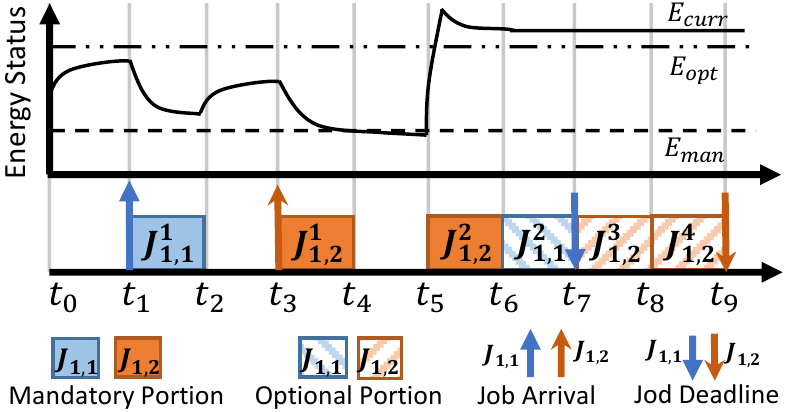}
\caption{Execution schedule of the workload.} 
\vspace{-15pt}
\label{fig:example}
\end{minipage}
\hspace{1em}
\begin{minipage}{0.48\textwidth}
\centering
\captionsetup{type=table} 
\vspace{5pt}
\caption{Explanation of the schedule in Figure~\ref{fig:example}.}
\vspace{-10pt}
\begin{footnotesize}
\begin{tabular}{|c|p{6.8cm}|}
\hline
\textbf{Time}                 & \multicolumn{1}{l|}{\textbf{Actions with Reasoning}}                            \\ \hline
$t_0$                  & There is no job in the system.                                              \\
$t_1$                  & $J_{1,1}^1$ (the only job) gets scheduled.                          \\
$t_2$                  & Since $E_{curr}$ < $E_{opt}$, optional $J_{1,1}^2$ is not scheduled.  \\
$t_3$                  & System prioritized $J_{1,2}^1$ over $J_{1,1}^2$ (See: Section~\ref{sec:scheduler}).     \\
$t_4$                  & Since $E_{curr}$ < $E_{man}$ no job is scheduled.                  \\
$t_5$                  & System prioritized mandatory $J_{1,2}^2$ over optional $J_{1,1}^2$ .  \\
$t_6$                  & Only optional jobs remain and $E_{curr}$ >$E_{opt}$. The system prioritizes $J_{1,1}^2$ over $J_{1,2}^3$ due to its tighter deadline. \\
$t_7$                  & $J_{1,2}^3$ (the only job) gets scheduled.                          \\
$t_8$                  & $J_{1,2}^4$ (the only job) gets scheduled.                          \\ \hline
\end{tabular}
\end{footnotesize}
\label{tab:example_des}

\end{minipage}
\vspace{-30pt}
\end{figure}

\subsection{Example Execution}
Table~\ref{tab:load_des} defines two jobs $J_{1,1}$ and $J_{1,2}$ from the same task, $\tau_1$. Figure~\ref{fig:example} demonstrates the execution of the jobs along with the status of harvested energy over a timeline. $E_{curr}$ refers to the current energy, and $E_{opt}$ and $E_{man}$ refer to two thresholds that determines whether the optional and the mandatory parts of a job should be executed, respectively. $J_{i,j}^k$ refers to the $k^{th}$ partition of job $J_{i,j}$. Table~\ref{tab:example_des} shows the action taken by the scheduler along with the reasoning at each time step.
Here, the mandatory part of $J_{1,2}$ is longer than the mandatory part  of $J_{1,1}$ because, for the second data sample, the classifier is not confident enough with the result after the execution of the first layer. Note that to simplify the illustration, we assume that each layer requires one time unit to execute and the partition (mandatory vs. optional) is known ahead of time. The proposed scheduling algorithm (described in Section~\ref{sec:scheduler}) handles further complexities such as the different execution times of different layers, multiple time units per layer, dynamic partitioning, and power-failure during the execution of a layer.

\parlabel{Setting $E_{man}$ and $E_{opt}$.}
The $E_{man}$ is set to the minimum energy required to turn on the MCU and execute an atomic fragment. 
During the compile time, \Sys programming tools (described in Section~\ref{sec:programming_model}) estimates the maximum energy required by any atomic fragment by running EnergyTrace++~\cite{EnergyTrace} and sets this threshold.  The $E_{opt}$, on the other hand, is by default set to the energy required to fill up the capacitor. This is because, once the capacitor is full, the excess energy gets wasted if nothing is executing on the MCU. By executing optional tasks, we minimize this wastage, and increase the performance of the system. However, a developer can override these values using the APIs provided (Section ~\ref{sec:apis_section}). If $E_{opt}$ is small, e.g., comparable or equal to $E_{man}$, then all the optional portion of the tasks will execute, causing starvation of the mandatory tasks. On the other hand, if $E_{opt}$ is high, the optional portion of the jobs will never execute.




\section{Modeling Intermittent Energy}
\label{sec:energy_study}

This section derives a single metric, namely the $\eta$-factor, which characterizes an energy harvester used in a particular application. The metric is later used in the real-time scheduler in \Sys.

\subsection{Energy Event}

The unavailability of harvestable energy and the required time to buffer sufficient energy before it can be used cause intermittence in batteryless systems. The pattern of intermittence is stochastic, hard to predict, and heavily dependent on the available harvestable energy. Hence, instead of modeling and predicting the harvested energy to characterize a harvester, we define a binary random variable, called the \emph{energy event}, that denotes the state of the energy storage to have at least $\Delta K$ Joules of energy over a $\Delta T$ period, where $\Delta K$ and $\Delta T$ depend on the application as well as the underlying system. In our empirical model, we set $\Delta K$ to $E_{man}$.



Furthermore, instead of directly dealing with the harvested energy, it is often easier to observe the physical phenomenon behind energy generation. For instance, for a piezoelectric harvester installed inside a smart shoe, the number of footsteps that generate at least $\Delta K$ Joules over $\Delta T$ time can be equivalently used to define an energy event. Likewise, a certain light-intensity for a solar harvester, and a certain number of packet transmissions for an RF harvester over $\Delta T$ time can be used to define energy events for these systems.


To characterize an energy harvester using the definition of energy events, we conduct a two-month long study on solar and RF harvesters (using empirically collected data) and human footsteps (using the dataset~\cite{meyer2016daily}) to analyze the pattern of energy events. Our study reveals that energy events occur in bursts, i.e., every harvester has a tendency to maintain its current binary state, and there is a probabilistic relation between consecutive energy events over a period. 

For instance, when a person starts walking, the probability that they will continue to walk is high over the next few time units and the probability decreases with time. Conversely, when a person is \emph{not} walking, the probability that they will continue not to walk is going to be high over the next few time units and will diminish with time. This observation enables us to impose conditional probability on future energy events, given the recent history of energy events---which is the key to characterize an energy harvester. We denote an energy event at time $t$ using the random variable $H_t \in \{0, 1\}$.



\subsection{Conditional Energy Event}
We define conditional energy event, $h(N)$ as the probability that an energy event will occur, given the immediately preceding $N$ consecutive energy events have occurred (for $N >0$) or have not occurred (for $N < 0$):


\begin{footnotesize}
\begin{equation}
    h(N) = 
    \begin{cases}
        p(H_t = 1 \mid H_{t-1} \wedge \ldots \wedge H_{t-N} = 1), & \text{for } N>0\\
        p(H_t = 1 \mid H_{t-1} \vee \ldots \vee H_{t-N} = 0), & \text{for } N<0
    \end{cases}
\end{equation}
\end{footnotesize}

To illustrate, h(10) = 90\% implies that an energy event will occur with 90\% probability if 10 immediately preceding consecutive energy events have occurred. 



\begin{figure}[!htb]
    \centering
    \vspace{-15pt}
    \subfloat[Ideal Source ($\eta$=1)]{\includegraphics[height=.13\textwidth]{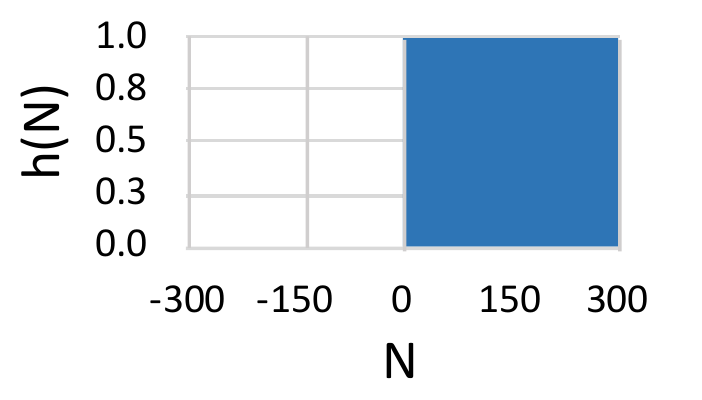}}
    \hspace{.1in}
    \subfloat[Footsteps ($\eta$=0.65)]{\includegraphics[height=.13\textwidth]{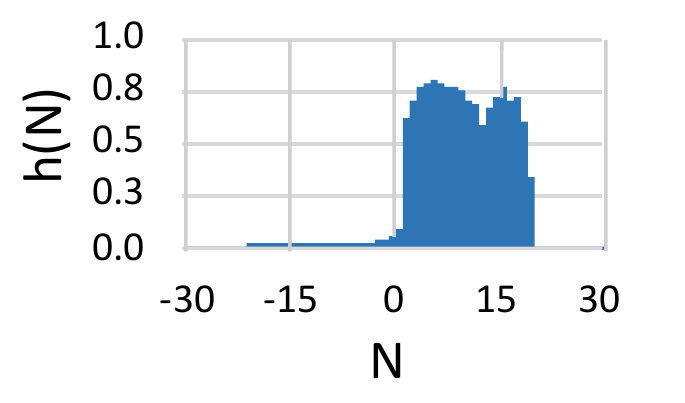}}
    \hspace{.1in}
    \subfloat[Stationary Solar ($\eta$=0.84)]{\includegraphics[height=.13\textwidth]{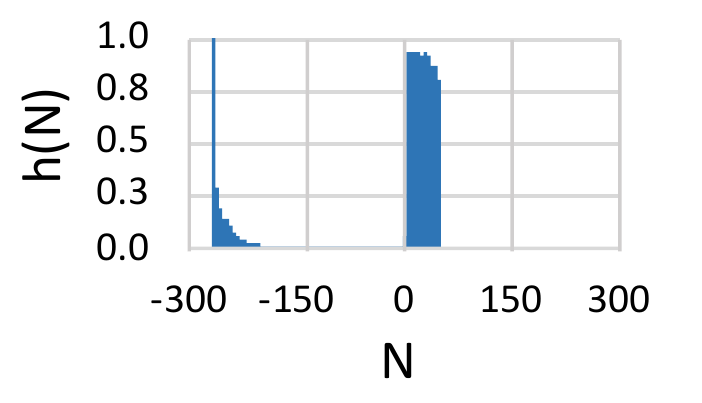}}
    \hspace{.1in}
    \subfloat[Stationary RF ($\eta$=0.76)]{\includegraphics[height=.13\textwidth]{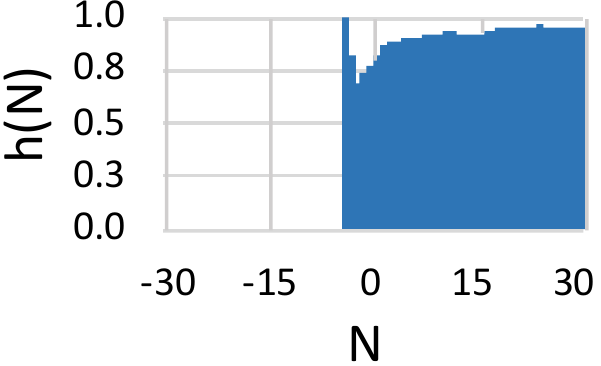}}
    \vspace{-5pt}
    \caption{Conditional energy event for (a) persistent power source, (b) piezo-electric harvester, (c) stationary solar harvester, and (d) stationary RF harvester. We use $\Delta T = 5$ minutes.}
    \label{fig:assumption_proof}
    \vspace{-1.5em}
\end{figure}




Figures~\ref{fig:assumption_proof}(a)-(d) show the distribution of $h(N)$ for a persistently-powered and three energy harvested systems. To characterize an energy harvester, we measure the Kantorovich-Wasserstein (KW) distance~\cite{kantorovich1942translation} between its distribution, $\mathscr{H}^{(i)}$ from an ideal (persistent power) source, $\mathscr{P}$ to obtain: 
\begin{footnotesize}
\begin{equation}
KW\big(\mathscr{H}^{(i)}, \mathscr{P}\big) = \int_{-\infty}^{+\infty} |CDF(\mathscr{H}^{(i)}) - CDF(\mathscr{P})|   
\end{equation}
\end{footnotesize}
In Figure~\ref{fig:assumption_proof}, we observe that $h(N)$ drops when $|N|$ increases. For example, in Figure~\ref{fig:assumption_proof}(b), $h(N)$ drops after $N=20$ since the person we studied never walked for more than 100 minutes. Similarly, in Figure~\ref{fig:assumption_proof}(c), after about five hours of consecutive energy events (i.e., light intensity > 2730 lux), the probability of energy event drops as the stationary solar harvester was placed beside a window that does not get enough light after five hours. We also notice that after about 19 hours of absence of any energy event (i.e., light intensity < 2730 lux), the probability of the next energy event is high as the sun shows up again at the window. 




\subsection{The \texorpdfstring{$\eta$}\xspace \xspace Factor}


Despite being informative, the KW distance has a limitation that not all $h(N)$'s are estimated using the same number of instances. Hence, we normalize the KW score against a purely random harvesting pattern, $\mathscr{R}$ to obtain a revised metric, called the $\eta$-factor:
\begin{footnotesize}
\begin{equation}
\eta = 1 - \frac{KW\big(\mathscr{H}^{(i)}, \mathscr{P}\big)}{KW\big(\mathscr{R}, \mathscr{P}\big)}
\end{equation} 
\end{footnotesize}
The value of $\eta$ lies in $[0, 1]$ and it measures how close a harvester's harvesting pattern is to a constant energy source. For a persistently-powered system, $\eta=1$, and for an energy harvester that shows no apparent pattern has $\eta=0$. For any other energy harvesting system, the $\eta$-factor will lie in-between and it is generally high for small $|N|$. A higher $\eta$-factor indicates less randomness in its energy harvesting pattern, and thus encourages a scheduler to make more aggressive decisions on scheduling tasks in the next few time slots. The $\eta$-factor needs to be empirically estimated for a given application-specific system. A critical discussion of $\eta$-factor is presented in Section~\ref{sec:discussion}.  


\section{Modeling DNN Tasks}
\label{sec:dnn_modeling}

In this section, we describe the task model of \Sys, training procedure of agile DNN, and construction of semi-supervised $k$-means classifier.

\subsection{Task Model}
\label{sec:dnn}




\parlabel{Tasks, Jobs, Units, and Fragments.} \Sys considers the processing of a sensor data stream for each classification task as an \emph{imprecise sporadic task}~\cite{liu1991algorithms}, $\tau_i = (T_i, D_i, C_i)$, where $T_i$ denotes the period (i.e., the minimum separation between two consecutive jobs), $D_i$ is the relative deadline, and $C_i$ is the worst-case execution time. 

An instance of a task, aka a \emph{job}, comprises of an ordered sequence of modules (henceforth these modules are called \emph{unit}s). The first $M$ units are mandatory and must be completed before the deadline, whereas the rest of the units of a job are optional and can be executed if time and resources are available. Such a partitioning scheme is known as the imprecise computing model in real-time systems literature~\cite{liu1991algorithms, zhang2004feedback, moallemi2011devs}. In this paper, however, the partition (i.e., the value of $M$) is dynamic and depends on the input data.    

In \Sys, a job that executes an $L$-layer agile DNN has $L$ units, where each unit corresponds to processing one DNN layer, along with the execution of the corresponding semi-supervised $k$-means classifier. The cluster centroids of this semi-supervised $k$-means classifier are updated with new unlabeled data. Based on the input data, \Sys may decide to exit from a unit or continue executing the next unit. The decision is based on a \emph{utility} function, which is described next.    

Although a \emph{unit} represents a logical grouping of related modules at each layer of the DNN, the size of a typical unit is generally too large to execute without an intermittence. Hence, at the implementation level, to avoid corrupted results and to ensure forward progress of code execution, these units are further divided into atomically executable \emph{fragments}---which guarantees correct intermittent execution using SONIC API~\cite{gobieski2019intelligence}.  





\begin{figure}[t]
\begin{minipage}{0.42\textwidth}
\centering
\vspace{2em}
\includegraphics[width=\textwidth]{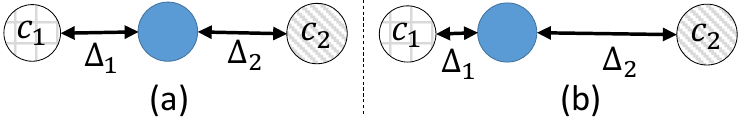}
\caption{Two nearest clusters $c_1$ and $c_2$ of an input (in the middle) is shown: (a) early exit does not happen since $|\Delta_2 - \Delta_1|$ is small; (b) early exit happens since $|\Delta_2 - \Delta_1|$ is large.}
\label{fig:policy}
\end{minipage}
\hspace{1em}
\begin{minipage}{0.44\textwidth}
\centering
\includegraphics[width=\textwidth]{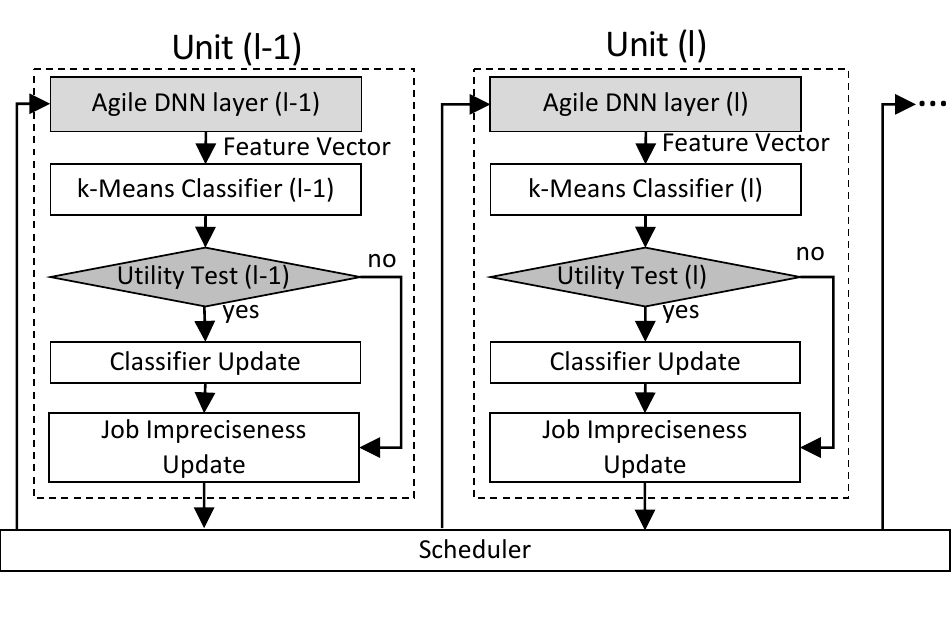}
\vspace{-3em}
\caption{Sequential execution of units of an agile DNN.}
\label{fig:dnn_flow}
\end{minipage}
\vspace{-2em}
\end{figure}

\parlabel{Dynamic Partitioning and Utility.}
Unlike traditional imprecise computing models~\cite{liu1991algorithms, zhang2004feedback, moallemi2011devs} where the partition of a job into mandatory and optional parts is known, the number of mandatory units in \Sys is determined at runtime. We propose a \emph{utility} function that estimates the confidence in classification at a given unit for that job. This represents the \emph{utility of the data} where higher utility at earlier units is desirable. 

Since we use a $k$-means classifier, we assume that a classification result is more likely to be correct if the input data sample being processed is unambiguously close to exactly one of the $k$ means. To achieve this, we compute the L1 distance of an input data sample (represented in terms a DNN layer and then vectorized) from two of its closest of the $k$ means, $\Delta_1$ and $\Delta_2$, and if their difference $|\Delta_2 - \Delta_1|$ is above a \emph{unit-specific} threshold, we decide to classify it as belonging to its closest cluster; otherwise, the computation of the DNN continues to the next unit. The process is illustrated by Figure~\ref{fig:policy}. 

The utility function described above runs in linear time with the number of clusters, i.e., $\mathcal{O}(k)$. It is lightweight, energy-efficient, and suitable for resource-constrained systems as it uses the byproduct of clustering-based classification which computes the cluster distances, $\Delta_i$'s anyways. It, however, depends on an offline-estimated threshold.   Section~\ref{sec:cluster_construction} describes how the utility threshold is computed using an empirical dataset. Section~\ref{sec:discussion} further discusses alternative utility functions that are suitable for other types of classifiers.

\parlabel{Preemption and Task Switching} \Sys allows limited preemption~\cite{baruah2005limited} where a job can be preempted only after a unit completes its execution. The scheduler kicks in at the completion of a unit and at the deadline of a job. After the execution of a unit, a job returns to the job queue with updated utility and imprecise status (mandatory or optional). Then the scheduler chooses the next highest priority job from the job queue using the priority function described in Section~\ref{sec:scheduler}. By prohibiting preemption of a unit, \Sys reduces context switching and read-write overheads, and minimizes the memory requirements to $\mathcal{O}(N)$ for $N$ jobs by using double-buffering ~\cite{asaro2000method}. Figure~\ref{fig:dnn_flow} shows the execution of two units. Each unit is shown as a large dotted rectangle and it contains four logical modules that are shown as solid rectangular boxes.



\subsection{Agile DNN Construction}
\label{sec:agile_cons}

Unlike previous works where inference happens only at the last layer~\cite{yang2016towards} or where a second classifier is used at hidden layers to decide early exit~\cite{teerapittayanon2016branchynet}, in \Sys, the output of \emph{any} hidden layer can be directly used as a feature that gets classified by a k-means classifier. Features obtained in this manner neither guarantee that the data samples from the same class are closer nor guarantee that the data samples from different classes are farther in the feature space. Hence, to ensure that the feature representation obtained after an early exit from the DNN execution maximizes the separability of different classes and minimizes the distance between examples of the same class, \Sys employs a \emph{layer-aware} loss function.



\parlabel{Layer-Aware Loss Function. } A convex combination of contrastive losses~\cite{islam2019soundsemantics} at each layer is used as the loss function of an Agile DNN -- which we call the \emph{layer-aware} loss function:
\begin{footnotesize}
\begin{align}
\label{eqn:eqlabelLA}
\begin{split}
     LA = \sum_{i=1}^{L} a_i \times LC\Big(W^i, X_1^i, X_2^i, \cdots , X_N^i\Big)
\end{split}
\end{align}
\end{footnotesize}
where, $a_i$ is the convex coefficient for the $i^{th}$ layer and $\sum_{i=1}^{L} a_i = 1$; $L$ and $N$ represent the total number of layers and classes, respectively; $W^i$ represent the weights of the $i^{th}$ layer; $X_1^i, X_2^i, \cdots , X_N^i$ are the output vectors corresponding to the members of each class at the $i^{th}$ layer; and $LC$ is the contrastive loss function. For two classes, $LC$ is defined as:
\begin{footnotesize}
\begin{align}
\label{eqn:eqlabelLC}
\begin{split}
    LC\Big(W^i, X_1^i, X_2^i\Big) = \dfrac{1}{2} \Big(1-Y\Big) \Big(G_{W^i}(X_1^i) - G_{W^i}(X_2^i)\Big)
    + \dfrac{1}{2} Y \max \Big(0, \Delta - G_{W^i}(X_1^i) - G_{W^l}(X_2^i)\Big)
\end{split}
\end{align}
\end{footnotesize}
where, $G_{W^i}(X_n^i)$ is the output feature set of a member of the $n$-th class ($1 \le n \le N$) at layer $i$. The coefficient $Y = 0$, if $X_1$ and $X_2$ belong to the same class, and $Y = 1$, otherwise. $\Delta$ represents the margin between the members of different classes in the feature space.

\begin{figure}[!htb]
\begin{minipage}{0.5\textwidth}
\centering
    \includegraphics[height=1in]{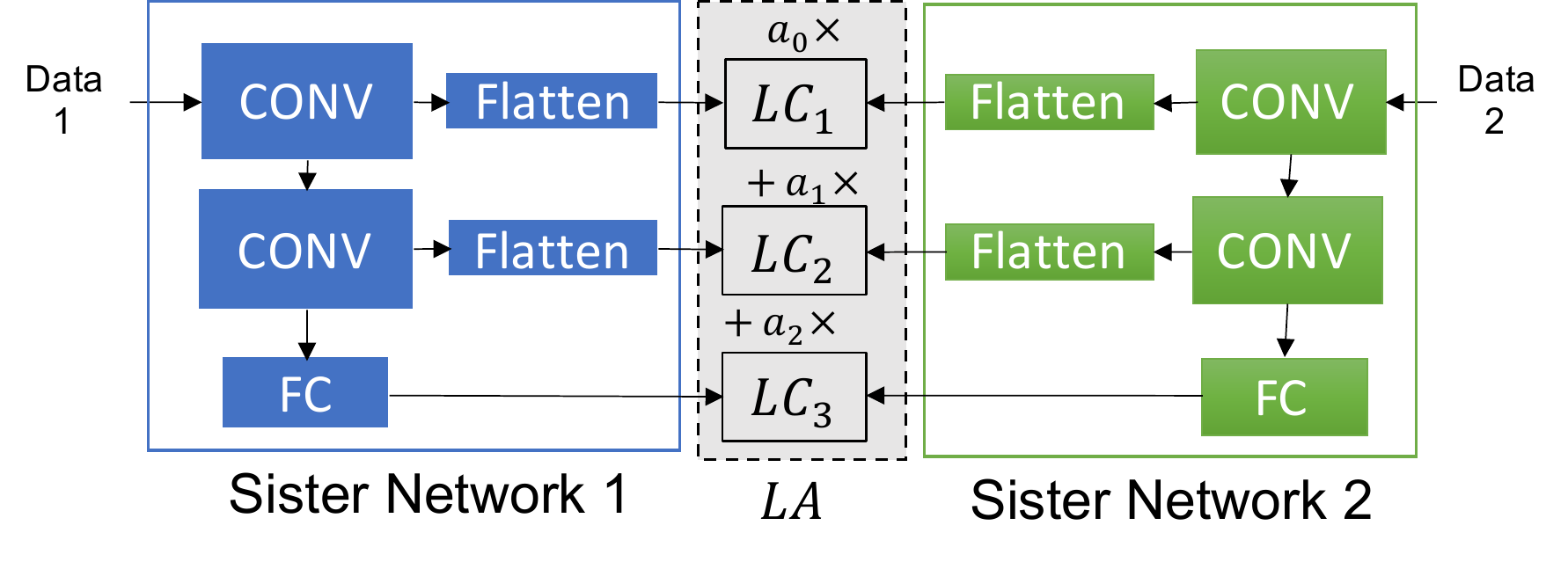}
    \vspace{-1em}
    \caption{Training agile DNN with Siamese network}
    \label{fig:siamese}
\end{minipage}
\hspace{1em}
\begin{minipage}{0.45\textwidth}
\centering
\includegraphics[height=1in]{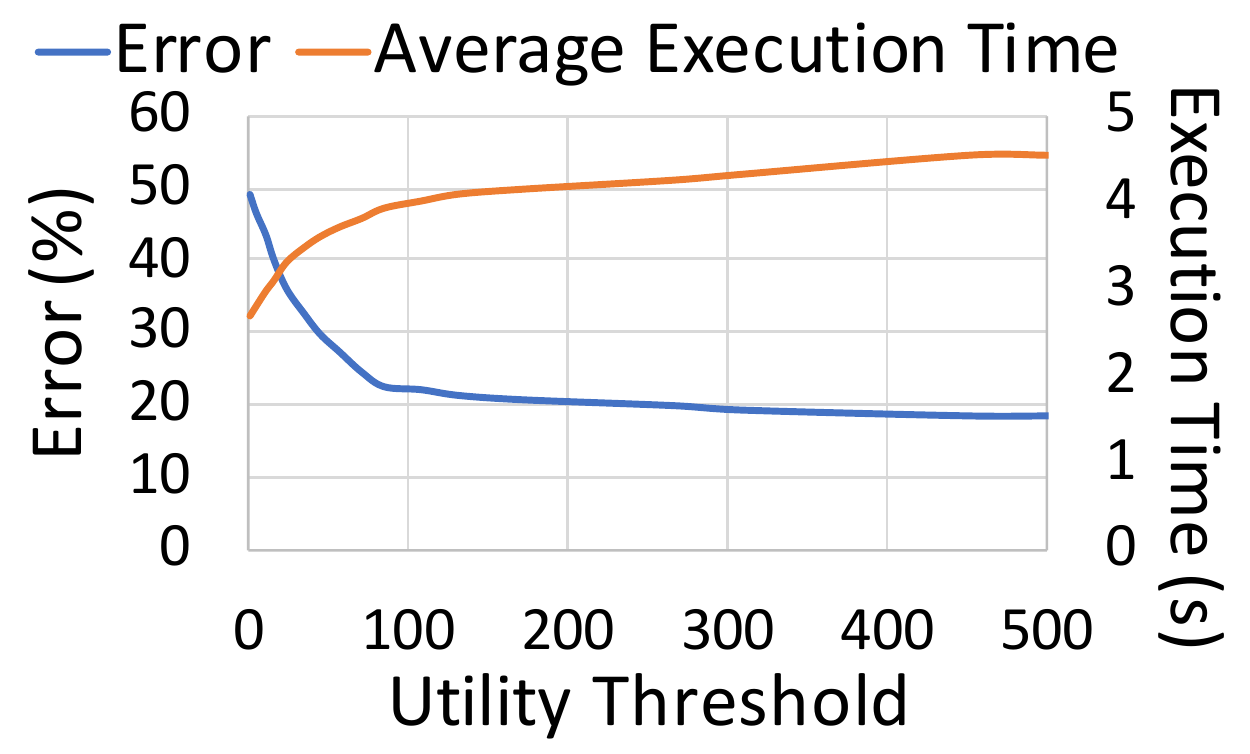}
\vspace{-1em}
\caption{Effect of utility threshold on performance}
\label{fig:utility_threshold}
\end{minipage}
\end{figure}

\parlabel{Training Agile DNN.}
To train an agile DNN, we use a siamese network architecture~\cite{islam2019soundsemantics} as shown in Figure~\ref{fig:siamese}. In a siamese network, there are two identical neural networks, called the sister networks, that share the same weights. From the labeled dataset, we select pairs of data points and use them as the inputs to the twin networks. Among the selected pairs of data points, 50\% belong to the same class, while the rest belong to different classes. Unlike~\cite{islam2019soundsemantics}, which only uses the contrastive loss at the last layer ($LC_3$), we use the layer-aware loss function (Equation~\ref{eqn:eqlabelLA}) at every layer to train these networks. We perform an exhaustive search for hyper-parameter tuning and to determine the weights of each layer. After the training, we use only one of the sister networks for inference. During inference, we obtain a representation of the input data from each layer, and use them as features for the semi-supervised $k$-means classifiers. Note that, for convolution layers, we flatten the output of a layer to get a vector instead of a tensor in order to be able to compute the L1-norm during the clustering-based classification step of \Sys.


Note that although the combination of an agile DNN and semi-supervised $k$-means classifiers in \Sys is inspired by \emph{Anytime Neural Networks}, an agile DNN is different as it is a representation learner rather than a classifier. Besides, an agile DNN is trained using a siamese network and it only has one loss function, which is different from anytime neural networks that use multiple auxiliary loss functions. Furthermore, the exit policy and the utility function of an agile DNN is different from that of anytime neural networks, and are optimized for resource-constrained systems. Moreover, \Sys forms an imprecise computing problem where an early exit from the network depends not only on the data but also on the time and energy budget.

\subsection{Semi-Supervised $k$-Means Classifiers Construction}
\label{sec:cluster_construction}
In \Sys, we maintain a semi-supervised $k$-means classifier at each unit. In this section, we describe how we construct and update these classifiers.

\parlabel{Computing Cluster Centroids.}
Using the trained agile DNN, we obtain a feature representation from each layer for each data point in the training dataset. Using these representations, we train a semi-supervised k-means classifier corresponding to each layer of the agile DNN (see Figure~\ref{fig:dnn_flow}). Using the labeled training data, we select the top N features using SelectKBest~\cite{george2017study} and $\chi^2$ tests, so that the features are computable on the resource-constrained target device. We utilize the labeled training data to determine the value of $k$ (for k-means) and assign a class label to each cluster. Finally, we compute the centroid of each cluster in the selected feature space.

\parlabel{Determining Utility Threshold.}
Utility thresholds are crucial to determining whether a data point should exit from the current hidden layer or continue processing through the network. A smaller threshold is likely to force too early exits and thereby, a lower classification accuracy; whereas a larger threshold is like to delay exits and thereby, increase the inference latency. This trade off is demonstrated by Figure~\ref{fig:utility_threshold} for the first layer of the DNN on the CIFAR-100 dataset. For different layers, we see similar trade offs. To determine a suitable utility threshold for each layer, we generate such a trade off curve and pick a utility threshold that ensures a desired minimum inference accuracy as configured by the programmer. 



\parlabel{Updating Centroids at Run-Time.}
We incrementally update the \emph{means}, i.e., the cluster centroids, of the k-means classifiers at runtime to evolve the classifiers over time and to learn from new examples---which is common in semi-supervised learning approaches~\cite{dhillon2002iterative,cheng2008constrained, basu2002semi}. Referring to Figure~\ref{fig:dnn_flow}, this is done inside the \emph{Classifier Adapter} when the classification result from the \emph{$k$-means Classifier} passes the \emph{Utility Test} at a unit. A new cluster centroid is computed by taking the weighted average of the current cluster centroid and the current example. Taking the weighted average guards against abrupt changes to the centroids due to the presence of an outlier or incorrect classifications. If the distribution of the input data points changes (e.g., the system is deployed in a new environment), the cluster centroid gradually shifts towards the new mean of the data points as it encounters the new data points.


\parlabel{Updating Centroids beyond Mandatory Layers.}
Due to early exit from the network, a data sample fails to update the k-means models of the deeper layers. To achieve this, \Sys adapts the cluster centroids of the deeper layers using the corresponding cluster heads of the layer from which the example exits early. Mathematically, the update operation is $c^{i+1} = \frac{1}{r}\sigma\big(W^{i+1} \times r \times c^i\big)$.
Here, $c^i$ and $c^{i+1}$ denote the corresponding cluster centroid of the k-means classifiers of layers $i$ and $i+1$; $W^{i+1}$ denotes weights (including the bias term) for layer $i+1$; $r$ denotes the size of a cluster; and $\sigma(x) = \frac{x + |x|}{2}$ is the non-linear activation function~\cite{schmidhuber2015deep}.

Since this technique estimates the cluster centroids of a deeper layer instead of actually running the data samples through those layers, it saves $\mathcal{O}(r)$ multiplication operations and performs the operation in $\mathcal{O}(1)$; at the maximum approximation error of $(\sum_{k=1}^r |W^{i+1} \times X_k^i| - |W^{i+1} \times \sum_{k=1}^r X_k^i|)/(2r)$.

\section{Real-Time Scheduler}
\label{sec:scheduler}

This section describes the real-time scheduler in \Sys. First, we introduce an online scheduling algorithm for dynamically-partitioned, sporadic, imprecise tasks on a persistently-powered system. Then, we extend the algorithm for intermittently-powered systems.

\subsection{Scheduler for Persistent Systems}
\label{sec:persistent}

Despite being an optimal online scheduling algorithm for sporadic tasks, the earliest deadline first (EDF) algorithm~\cite{krishna2001real} is not directly applicable to \Sys as EDF does not consider the accuracy of a DNN. Furthermore, traditional scheduling algorithms for imprecise tasks~\cite{liu1991algorithms, shi2009hash} are not directly applicable to \Sys as well since the mandatory and optional portions of an agile DNN is determined dynamically at runtime. 

To address these challenges, we propose a priority function to prioritize the \emph{units} of \Sys. At the end of the execution of an unit, the scheduler selects the highest priority unit as the next unit for execution. The priority function considers not only the remaining deadline of a job, but also the utility (as defined in Section~\ref{sec:dnn}) and the dynamically determined impreciseness status (i.e., mandatory vs. optional) of a unit:

\begin{footnotesize}
\begin{equation}
    \zeta_{i,j}^l = \Big(1 - \alpha (d_{i,j} - t_c)\Big) + \Big(1 - \beta \Psi_{i,j}^l\Big) + \gamma_{i,j}^l
\end{equation}
\end{footnotesize}
where the first term represents the remaining deadline, which is the difference between a job's absolute deadline $d_{i,j}$ and the current time $t_c$. The second term ensures that units with lower utility score $\Psi_{i,j}^l$ gets higher priority as these tasks need further execution for accurate classification. The third term is a binary variable $\gamma_{i,j}^l \in \{0, 1\}$ that denotes if the unit under consideration is mandatory $\gamma_{i,j}^l = 1$ or optional $\gamma_{i,j}^l = 0$, which is determined at runtime based on the unit-specific utility threshold. $\alpha$ and $\beta$ are scaling parameters that normalize the deadline and utility, which are the inverse of the maximum deadline and utility, respectively.

Note that, \Sys supports multiple tasks including multiple DNN tasks as long as the required memory does not exceed the available memory of the system. For non-DNN tasks or other absolute (non-imprecise) tasks, $\gamma_{i}$ is always 1 and $\Psi_{i}$ is a constant for all units. $\Psi_{i}$ is user-defined based on the priority of the task.

\subsection{Scheduling for Intermittent System}
\label{sec:intermittent}

For an intermittently-powered system, we utilize the $\eta$-factor introduced in Section~\ref{sec:energy_study} to extend $\zeta$ as follows:
\begin{footnotesize}
\begin{equation}
    \zeta_{I_{i,j}}^l = 
    \begin{cases}
    \big(1 - \alpha (d_{i,j} - t_c)\big) + \big(1 - \beta \Psi_{i,j}^l\big) + \gamma_{i,j}^l, &  \eta E_{curr} \geq E_{opt}\\
    \gamma_{i,j}^l \Big( \big(1 - \alpha (d_{i,j} - t_c)\big) + (1 - \beta \Psi_{i,j}^l) \Big), &  \eta E_{curr} < E_{opt}
    \end{cases}
\end{equation}
\end{footnotesize}

Here, $E_{curr}$ is the current energy of the system and $E_{opt}$ is a threshold that determines if the system has enough energy to execute both mandatory and optional units. The expression $\eta E_{curr}$ is high enough to cross the threshold as long as at least one of the two variables $\eta$ and $E_{curr}$ is high-valued and the other is not extremely low. We identify two cases: 

First, when $\eta E_{curr}$ is above the threshold, both mandatory and optional units are considered for scheduling. Intuitively, it captures the cases when (a) an energy harvester is predictable and generating at least sufficient energy to keep the capacitor charged, and (b) when an energy harvester is predictable with medium confidence and generating more than sufficient energy.
We omit the explanation of the three terms in this case since they are similar to the persistent power system as described in the previous section. 

Second, when $\eta E_{curr}$ is below the threshold, only the mandatory units are considered for scheduling. It captures the cases when an energy harvester is -- (a) unpredictable, (b) predictable but generates insufficient energy, and (c) predictable with medium confidence and generates sufficient energy. 

$\zeta_I$ minimizes two types of energy waste in batteryless systems: 1) wasted energy due to executing unnecessary portions of a job, and 2) wasted energy due to not executing any job while the harvester gets enough energy from the source to keep the capacitor charged~\cite{buettner2011dewdrop}. The first type of waste is avoided by scheduling conservatively when $\eta E_{curr} < E_{opt}$, and the second type of waste is avoided by executing optional units when $\eta E_{curr} \geq E_{opt}$.



\subsection{Schedulability Condition} 
\label{sec:condition}

A set of $N$ sporadic tasks is schedulable by an imprecise scheduler if the total utilization, $\sum_{i=1}^N \frac{C_i}{T_i} \leq 1$~\cite{liu1991algorithms}, where the execution time, $C_i$ includes only the mandatory portion of the task. Scheduling sporadic jobs in an intermittently-powered system adds further complexity as power outages essentially blocks the CPU and thus increases the CPU utilization by increasing the execution time $C_i$, although no task is actually executing on the system as power is out. In order to incorporate energy intermittence into the schedulability analysis framework, we model power outages event as a very high-priority job of a sporadic \emph{Energy Task}. 



In this extended task set having $N+1$ tasks, the schedulability condition becomes $\sum_{i=1}^N \frac{C_i}{T_i} + \frac{C_e}{T_e} \leq 1$, where, $C_e$ and $T_e$ are the duration and interval of energy intermittence. The execution time of an energy task, $C_e$ is related to the $\eta$-factor of the system. The probability that an energy harvester will remain in its current power-outage state for the next $d$ energy events can be derived from $\eta$ using the properties of a geometric distribution $\eta^d (1-\eta)$, whose expected value is $E[C_e] = \eta /(1-\eta)$. Given this, the necessary condition for an intermittent computing system to be able to schedule $N$ sporadic tasks is $T_E \geq \frac{\eta/ (1-\eta)}{1 - \sum_{i=1}^N (C_i / T_i)}$.







\begin{figure}[!htb]
    \centering
    \includegraphics[width=\textwidth]{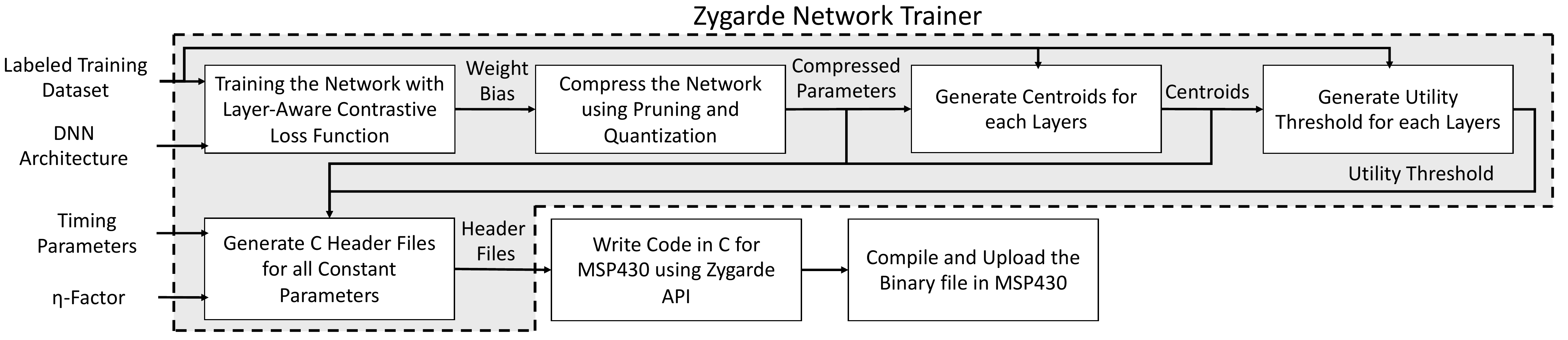}
    \vspace{-10pt}
    \caption{\Sys Programming Framework}
    \label{fig:my_label}
\end{figure}
\section{\Sys Programming Model}
\label{sec:programming_model}

\Sys's programming model consists of: (i) a \emph{Network Trainer} tool that is used by the developer to train and compress agile DNNs and to generate the $k$-means classifiers and corresponding hyper-parameters; and (ii) \emph{APIs} for the target embedded device which are used by the developer to write custom C application for an intermittently-powered MSP430 MCU. A high-end development machine is recommended for these one-time, offline steps. At the end of these steps, we obtain an executable binary file for the MSP430 MCU.

\subsection{\Sys Network Trainer}
\label{sec:network_trainer}

The network trainer takes four inputs from the developer, i.e., (1) a labeled training dataset, (2) DNN architecture/model, (3) timing parameters, and (4) the $\eta$-factor. The network trainer generates C header files as the output -- which are used by the APIs for the target embedded platform described in the next section. Figure~\ref{fig:my_label} shows the intermediate steps inside the \Sys network trainer.

At first, the network trainer trains the agile DNN model using the labeled dataset using the layer-aware loss function described in Section~\ref{sec:agile_cons}. It relies on an exhaustive search for hyper-parameter tuning, and outputs the weights and bias parameters of the network. Considering the limited memory of the target device, the DNN is compressed and pruned to reduce its memory requirement~\cite{bhattacharya2016sparsification, chollet2017xception, de2000multilinear, de2000best, tucker1966some, han2015deep, nabhan1994toward}. The network trainer also checks if the  compressed network fits into the memory of the target device and signals an error if it does not. Using this compressed agile DNN model and the input dataset, the network trainer generates the cluster centroids and the utility threshold for each layer of the network -- following the steps described in Section~\ref{sec:cluster_construction}.


Finally, C header files are generated that contain the compressed DNN parameters, cluster centroids, utility thresholds, features used in clustering, and task-specific timing parameters (e.g., deadline and period) and the energy parameter (i.e., $\eta$-Factor).

\subsection{\Sys APIs}
\label{sec:apis_section}

The \Sys APIs extend open source SONIC~\cite{gobieski2019intelligence} APIs for intermittent DNN computing by incorporating \Sys-specific capabilities, such as early termination, cluster-based inference, and scheduling. We divide the APIs into two categories: (1) external APIs, and (2) internal APIs.

\begin{figure}
\begin{minipage}{0.24\textwidth}
\centering
    \vspace{1em}
    \includegraphics[width=\textwidth]{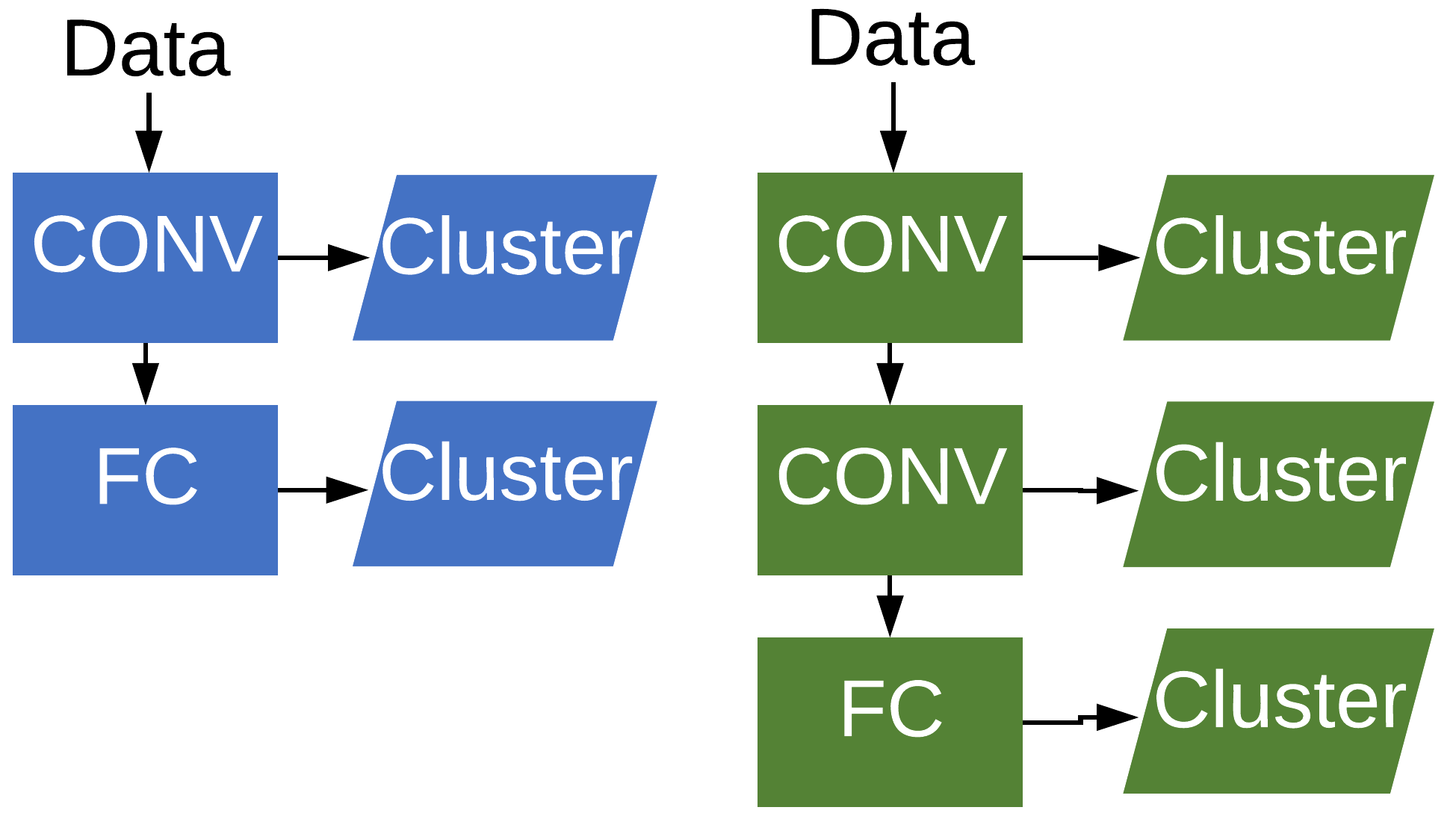}
    \caption{Two sample DNNs.}
    \label{fig:sampleDNN}
\end{minipage}
\begin{minipage}{0.65\textwidth}
    \centering
    \includegraphics[width=\textwidth]{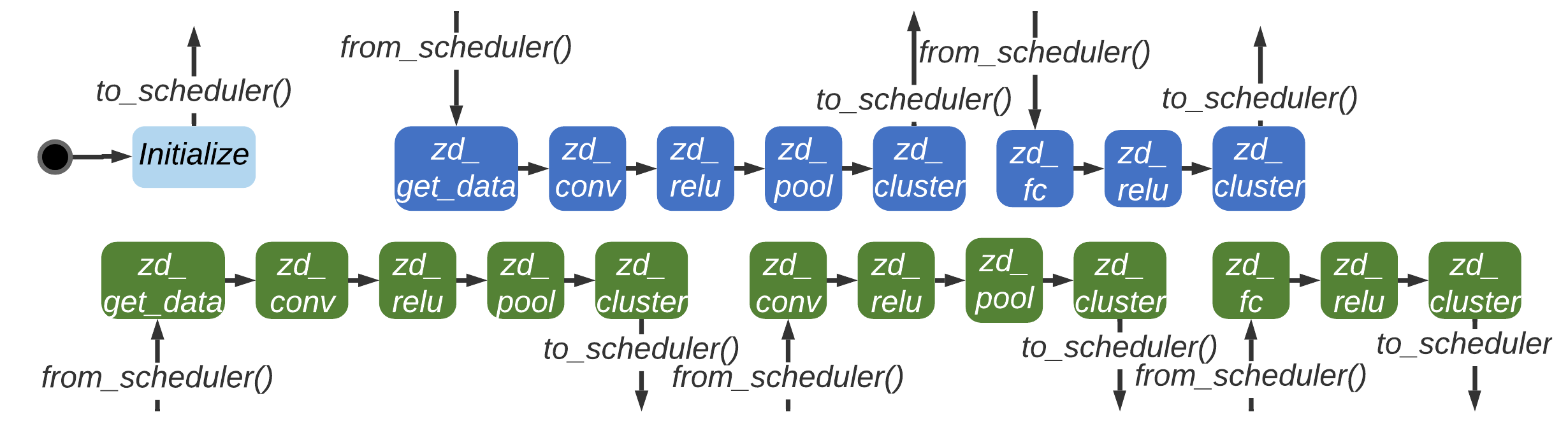}
    \vspace{-3em}
    \caption{State diagram of the sample DNNs.}
    \label{fig:sampleA}
\end{minipage}
\vspace{-1.7em}
\end{figure}

The external APIs contain library functions that a developer uses to implement early-exit capable agile DNNs for feature representation and the $k$-means classifiers. These library functions rely on the header files generated by the network trainer to access the classifier parameters and are sufficient for most developers who only want to define the high-level logic of their application. For instance, to implement the two tasks shown in Figure~\ref{fig:sampleDNN} on an MSP430 platform, a developer essentially has to write a C program that uses \Sys external APIs to implement a state diagram similar to the one shown in Figure~\ref{fig:sampleA}.

The internal APIs provide some of the lower level functions that are primarily used by the external APIs. These APIs implement several key features of \Sys including the scheduler, job queue management, time management, and handling the timers. 
If a developer wishes to change the default implementation of any of these functions, they need to override these methods to provide their own implementation.

\section{Implementation}
\label{sec:setup}
\begin{figure}[!htb]
     \centering
     \subfloat{
         \includegraphics[height=0.2\textwidth]{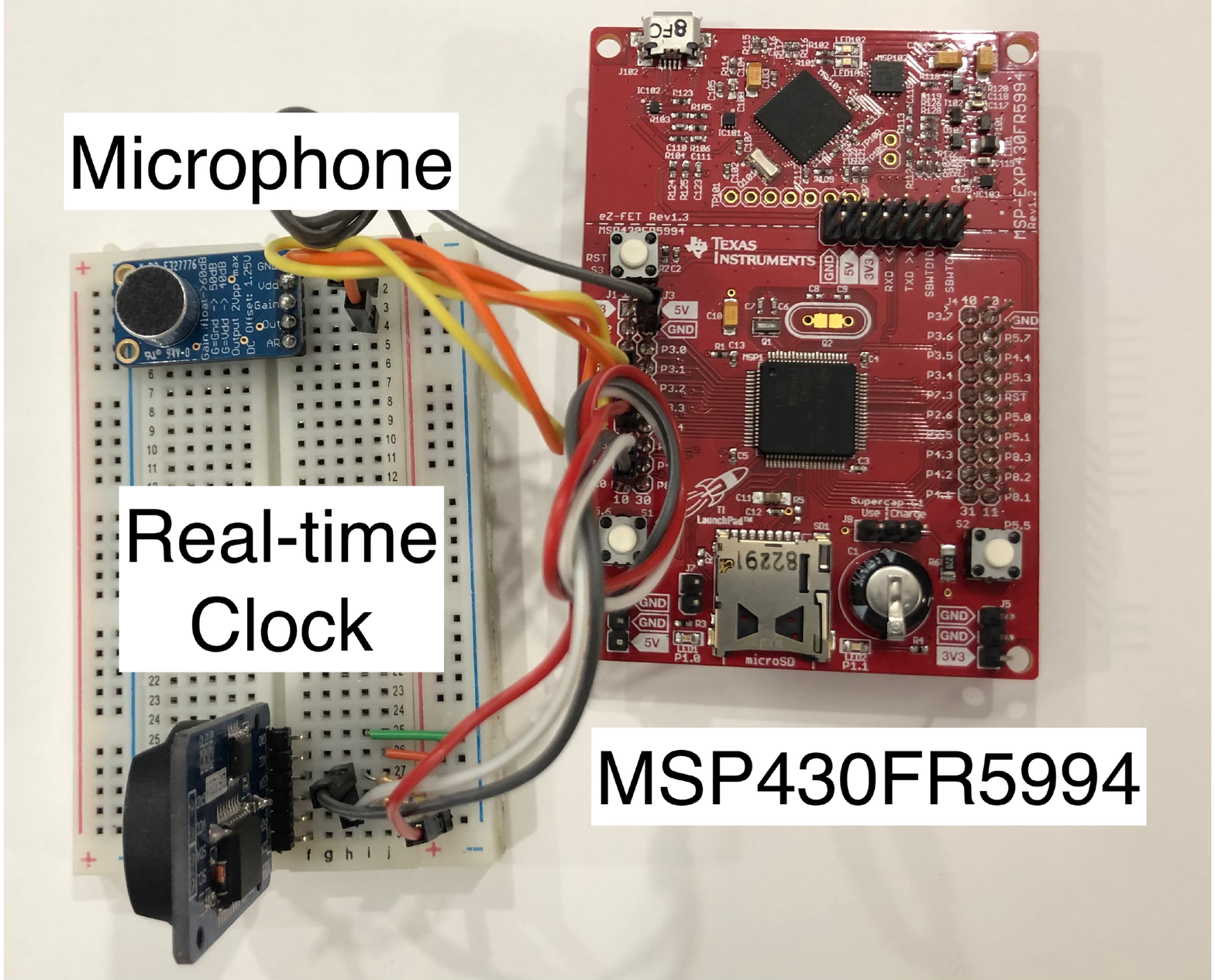}
     }
     \subfloat{
         \includegraphics[height=0.2\textwidth]{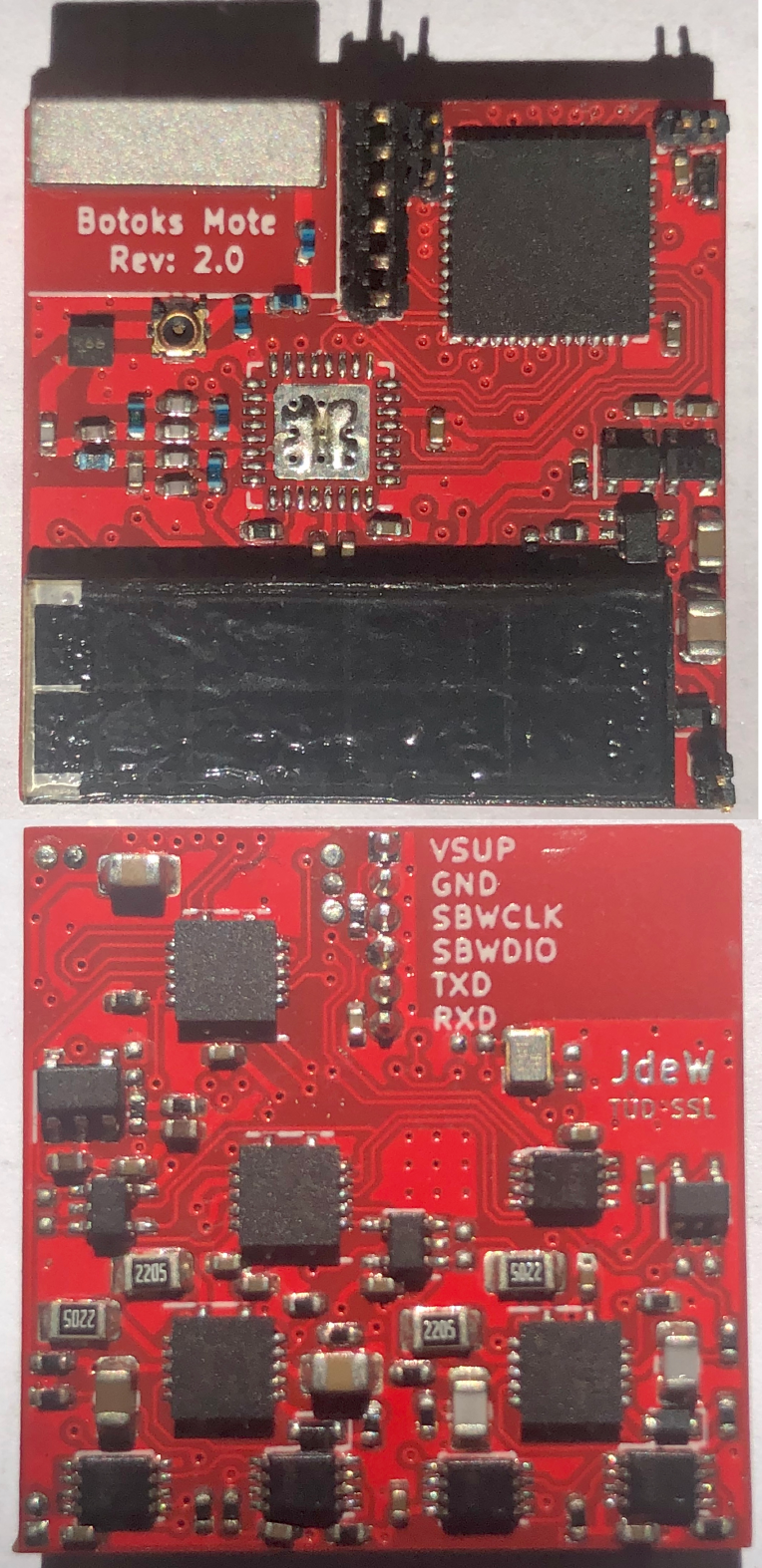}
     }
     \subfloat{
         \includegraphics[height=0.2\textwidth]{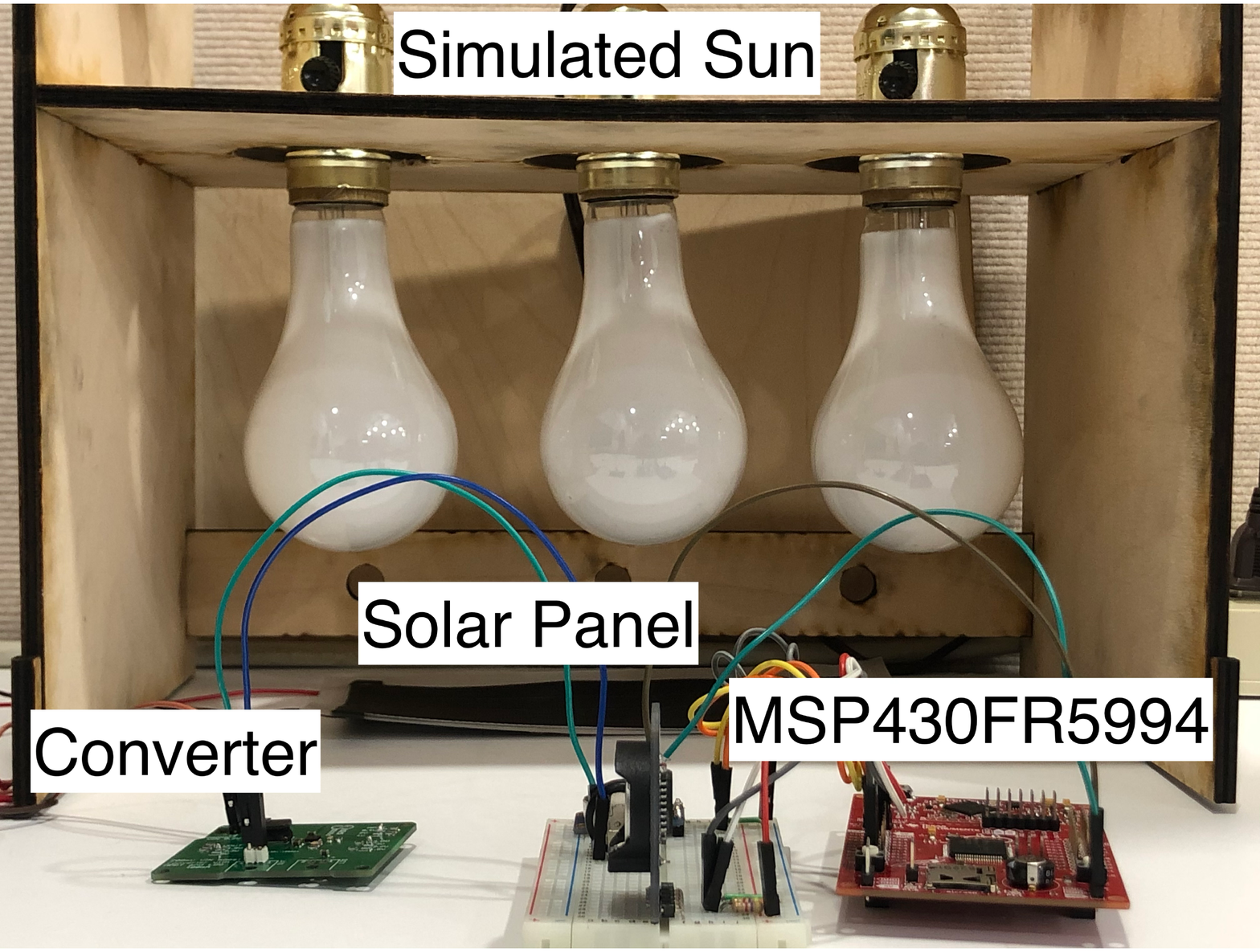}
     }
     \subfloat{
         \includegraphics[height=0.2\textwidth]{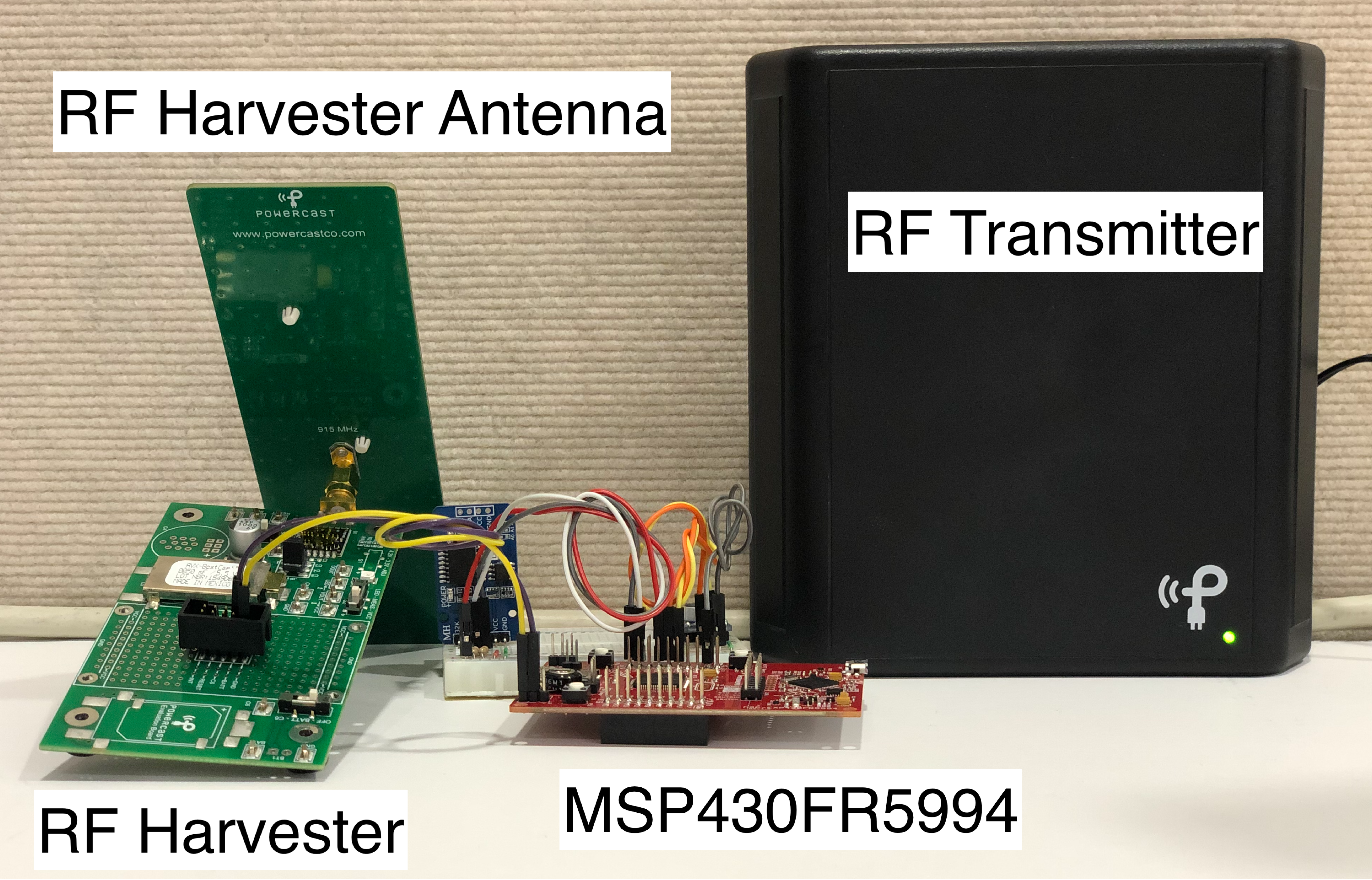}
     }
     \vspace{-0.5em}
    \caption{\Sys experimental setup. }
    \label{fig:hardware}
    \vspace{-1.8em}
\end{figure}

\parlabel{Computing Device.}
We use TI-MSP430FR5994~\cite{msp430} MCU (shown in Figure~\ref{fig:hardware}) that has 256KB of FRAM, 8KB of SRAM, 6-channel DMA, a low energy accelerator (LEA), and an operating voltage range of 1.8V to 3.6V. 
During the training phase, we use an Intel Core i7 PC with RTX2080 GPU to train and compress the agile DNN, initialize the centroids of semi-supervised $k$-means classifiers, and compute the utility thresholds.

\parlabel{Energy Harvester.}
Figure~\ref{fig:hardware} shows our solar and RF energy harvester setup. The solar harvester includes an Ethylene Tetrafluoroethylene (ETFE) based solar panel~\cite{solarpanel} and a step-up regulator~\cite{ltc3105}.
We use the Powercast harvester-transmitter pair~\cite{powercast, powercasetransmitter} to harvest RF energy. Like previous works on intermittent systems~\cite{gobieski2019intelligence, maeng2017alpaca}, both harvesters use a 50mF capacitor.

\parlabel{Sensor Peripheral.}
We use an electret microphone~\cite{BOB} and the built-in ADC in MSP430 for acoustic sensing. For visual sensing, we use an OV2640 CMOS camera module~\cite{camera} connected via I2C and SPI. Using LEA and DMA, we perform FFT on audio data and write the audio data to the FRAM without involving the CPU. 


\parlabel{Time Keeping.}
Like ~\cite{yildirim2018ink,hester2017timely}, we use a real-time clock, DS3231~\cite{timer} for timekeeping in most of the experiments. We use this clock only during the power up to sync and maintain the internal clocks of the MCU. This clock is easily replaceable with an SRAM or capacitor-based timekeeping system during power outages~\cite{rahmati2012tardis, hester2016persistent}. In order to quantify the effect of such a batteryless timekeeper on \Sys, we implement and use an open-source remanence clock, called CHRT~\cite{de2020reliable}, in one of the experiments. To use the CHRT correctly with \Sys, the energy required to charge the CHRT has been considered when defining the energy events to estimate the $\eta$-factor.


\parlabel{Libraries.}
Zygarde uses an open-source intermittent execution model SONIC~\cite{gobieski2019intelligence} and related APIs (e.g., ALPACA~\cite{maeng2017alpaca}). We use Tensorflow~\cite{45381} for training the DNN models.

\section{Microbenchmarks}
\label{sec:eval}
In this section, we evaluate each component of \Sys using datasets and compare \Sys with baseline algorithms. We also observe the effect of capacitor size and remanence clock on \Sys.

\subsection{Datasets and Environments}
\label{sec:dataset}
\parlabel{Datasets and DNNs.}
To evaluate the performance of different components of \Sys, we use four datasets: MNIST~\cite{lecun1998mnist}, ESC-10~\cite{piczak2015dataset}, CIFAR-100~\cite{krizhevsky2009learning}, and Visual Wake Word~\cite{chowdhery2019visual}. MNIST is a popular image dataset having 80,000 $28\times28$ pixel images (60,000 for training and 10,000 for testing) and ten classes, and it has been used for evaluating state-of-the-art intermittent computing systems~\cite{gobieski2019intelligence}. 
ESC-10 also has ten classes and 44.1 kHz five seconds-long audio clips. We use 1s audio downsampled to 8KHz. We split the dataset into 80\% training and 20\% testing datasets. CIFAR-100 contains $32\times32$ pixel color images from 100 classes. It has 500 training images and 100 testing images per class. In order to fit this dataset in the MSP430, we use randomized subsets of 5 classes from the dataset for 100 iterations and report the average. 

\begin{figure}[]
    \centering
    \subfloat[Original Image]{\includegraphics[height=.22\textwidth]{fig/original.jpg}}
    \hspace{.1in}
    \subfloat[Downsampled Only]{\includegraphics[height=.22\textwidth]{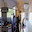}}
    \hspace{.1in}
    \subfloat[Targeted Crop]{\includegraphics[height=.22\textwidth]{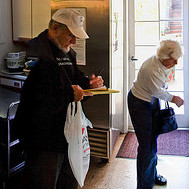}}
    \vspace{-1em}
    \caption{Visual Wake Word dataset: (a) Original image (640$\times$320), (b) Only downsampled (32$\times$32), (c) After targeted cropping and downsampling (32$\times$32).}
    \label{fig:vww_downsample}
    \vspace{-1em}
\end{figure}

Visual Wake Word (VWW) is a large dataset containing 82,783 training and 40,504 validation images from the state-of-the-art vision dataset COCO~\cite{lin2014microsoft}. To fit these images into the MCU’s memory, we first crop an image to move the target object (human) in the center and then downsample the cropped image to $32\times32$ pixels. Our dataset can be accessed at~\cite{vww_dataset}. Note that if we only downsample the image to $32\times32$ pixels without cropping it first, the resultant image scales down the target object (human) so much that they are not recognizable anymore. Figure~\ref{fig:vww_downsample} shows an example image from the VWW dataset, followed by two downsampled versions of it -- with and without cropping.



We implement four compressed networks summarized in Table~\ref{tab:network}. Our feature-maps after each layer consist of a maximum of 150 features selected using $k$-best select. These feature-maps are used for the semi-supervised $k$-means classifiers. The scheduler has a queue-size of 3.

\parlabel{Controlled Energy Sources.}
To evaluate the system with different $\eta$-factors ($\Delta T$=1s and $\Delta K$=9.36mJ), we perform controlled experiments. To determine the value of $\Delta K$, we run the system for multiple iterations and take the highest observed energy consumption. We vary the distance between the transmitter and the receiver between 1-5 feet for RF. We simulate solar power with three dimmable bulbs with varying intensity (5.6 Klx - 35 Klx) as shown in Figure~\ref{fig:hardware}. The seven scenarios considered for the evaluation is described in Table~\ref{tab:scenario}. Note that, we use outdoor scenarios and windowed rooms to get the sunlight for the real-life experiments in Section~\ref{sec:real_evaluation}.

\begin{figure}[!htb]
\begin{minipage}{.55\textwidth}
\captionsetup{type=table} 
\caption{DNNs considered in this section.}
\vspace{-10pt}
\begin{footnotesize}
\centering
\begin{tabular}{|p{.25\textwidth}|c|c|c|c|}
\hline
\textbf{Dataset} &
  \textbf{MNIST} &
  \textbf{ESC-10} &
  \textbf{CIFAR-100} &
  \textbf{VWW} \\ \hline
\textbf{Layers} &
\multicolumn{1}{c|}{\begin{tabular}[t]{@{}c@{}}CONV\\ CONV\\ FC\\ FC\end{tabular}} &
  \multicolumn{1}{c|}{\begin{tabular}[t]{@{}c@{}}CONV\\ CONV\\ CONV\\ FC\end{tabular}} &
  \multicolumn{1}{c|}{\begin{tabular}[t]{@{}c@{}}CONV\\ CONV\\ FC\\ FC\end{tabular}} &
  \multicolumn{1}{c|}{\begin{tabular}[t]{@{}c@{}}CONV\\ CONV\\ CONV\\ CONV\\ FC\end{tabular}} \\ \hline
\textbf{Dimensions} &
  \multicolumn{1}{c|}{\begin{tabular}[t]{@{}c@{}}20$\times$1$\times$5$\times$5\\ 100$\times$20$\times$5$\times$5\\ 200$\times$1600\\ 500$\times$200\end{tabular}} &
  \multicolumn{1}{c|}{\begin{tabular}[t]{@{}c@{}}16$\times$1$\times$5$\times$5\\ 32$\times$16$\times$5$\times$5\\ 64$\times$32$\times$5$\times$5\\ 95$\times$256\end{tabular}} &
  \multicolumn{1}{c|}{\begin{tabular}[t]{@{}c@{}}32$\times$3$\times$5$\times$5\\ 64$\times$32$\times$5$\times$5\\ 384$\times$1600\\ 192$\times$384\end{tabular}} &
  \multicolumn{1}{c|}{\begin{tabular}[t]{@{}c@{}}16$\times$3$\times$5$\times$5\\ 32$\times$16$\times$5$\times$5\\ 64$\times$32$\times$5$\times$5\\  64$\times$64$\times$5$\times$5\\ 192$\times$256\end{tabular}} \\ \hline
\textbf{Parameters Size} &
  $8\times{10}^3$ &
  $55\times{10}^3$ &
  $27\times{10}^3$ &
  $14\times{10}^3$ \\ \hline
\end{tabular}
\end{footnotesize}
\label{tab:network}
\end{minipage}
\hspace{2em}
\begin{minipage}{.37\textwidth}
\vspace{-27pt}
\captionsetup{type=table} 
\caption{Algorithm Evaluation Scenarios}
\vspace{-10pt}
\begin{footnotesize}
\centering
\begin{tabular}{|c|c|c|c|}
\hline
\textbf{System} & \begin{tabular}[c]{@{}c@{}}\textbf{Energy}\\ \textbf{Source}\end{tabular} & \textbf{$\eta$} & \begin{tabular}[c]{@{}c@{}}\textbf{Average}\\ \textbf{Power (mW)}\end{tabular} \\ \hline
1 & Battery & 1    &       \\ 
2 & Solar   & 0.71 & 600 \\ 
3 & Solar   & 0.51 & 420 \\ 
4 & Solar   & 0.38 & 310 \\ 
5 & RF      & 0.71 & 58  \\ 
6 & RF      & 0.51 & 71  \\ 
7 & RF      & 0.38 & 80  \\ \hline
\end{tabular}
\vspace{.5em}
\label{tab:scenario}
\end{footnotesize}
\end{minipage}
\vspace{-2em}
\end{figure}



\subsection{System Overhead}
\label{sec:overhead}
\begin{figure}[!htb]
    \centering
    \vspace{-1.5em}
    \subfloat{\includegraphics[height=.17\textwidth]{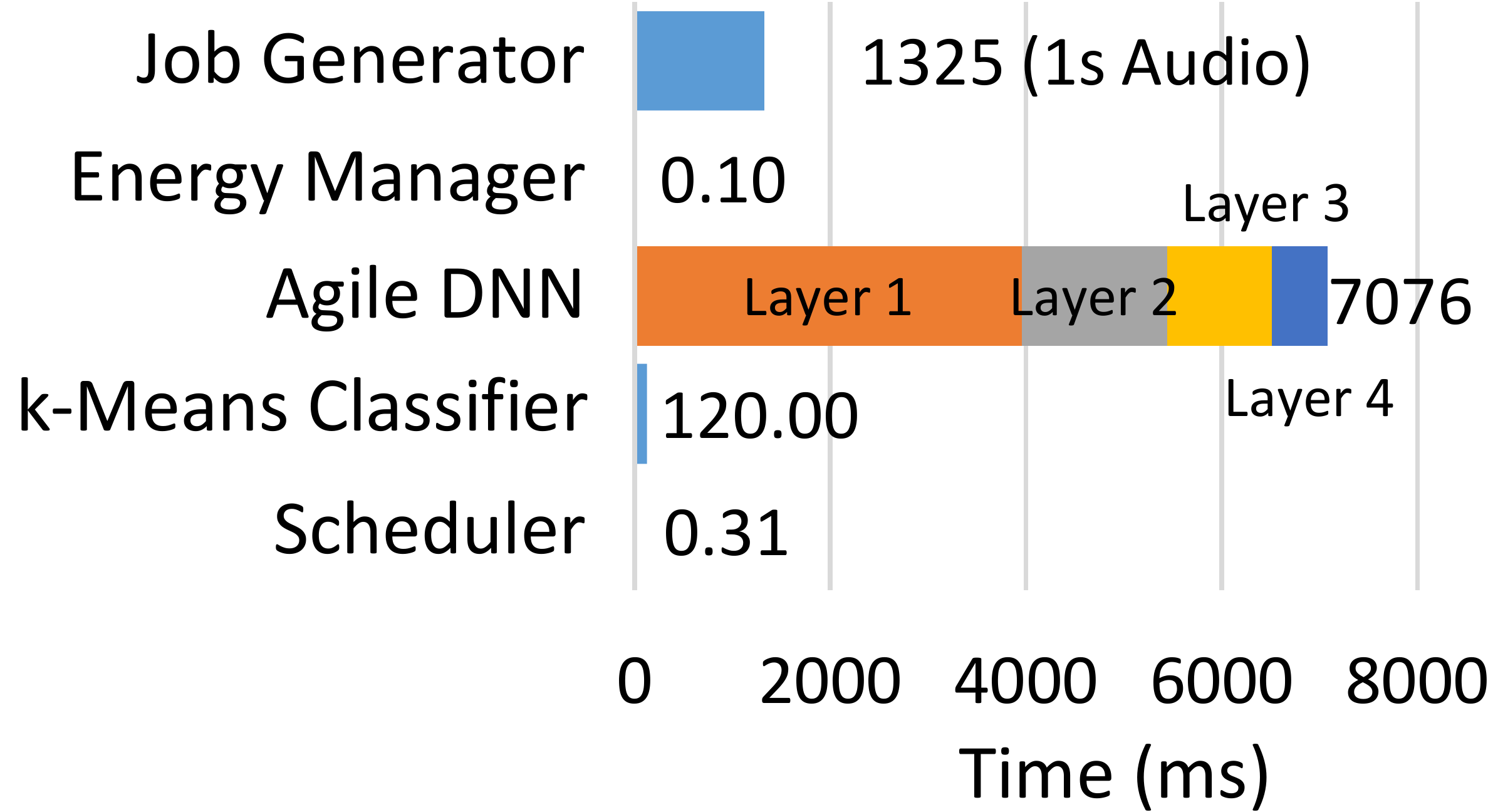}}
    \hspace*{.1\textwidth}
    \subfloat{\includegraphics[height=.17\textwidth]{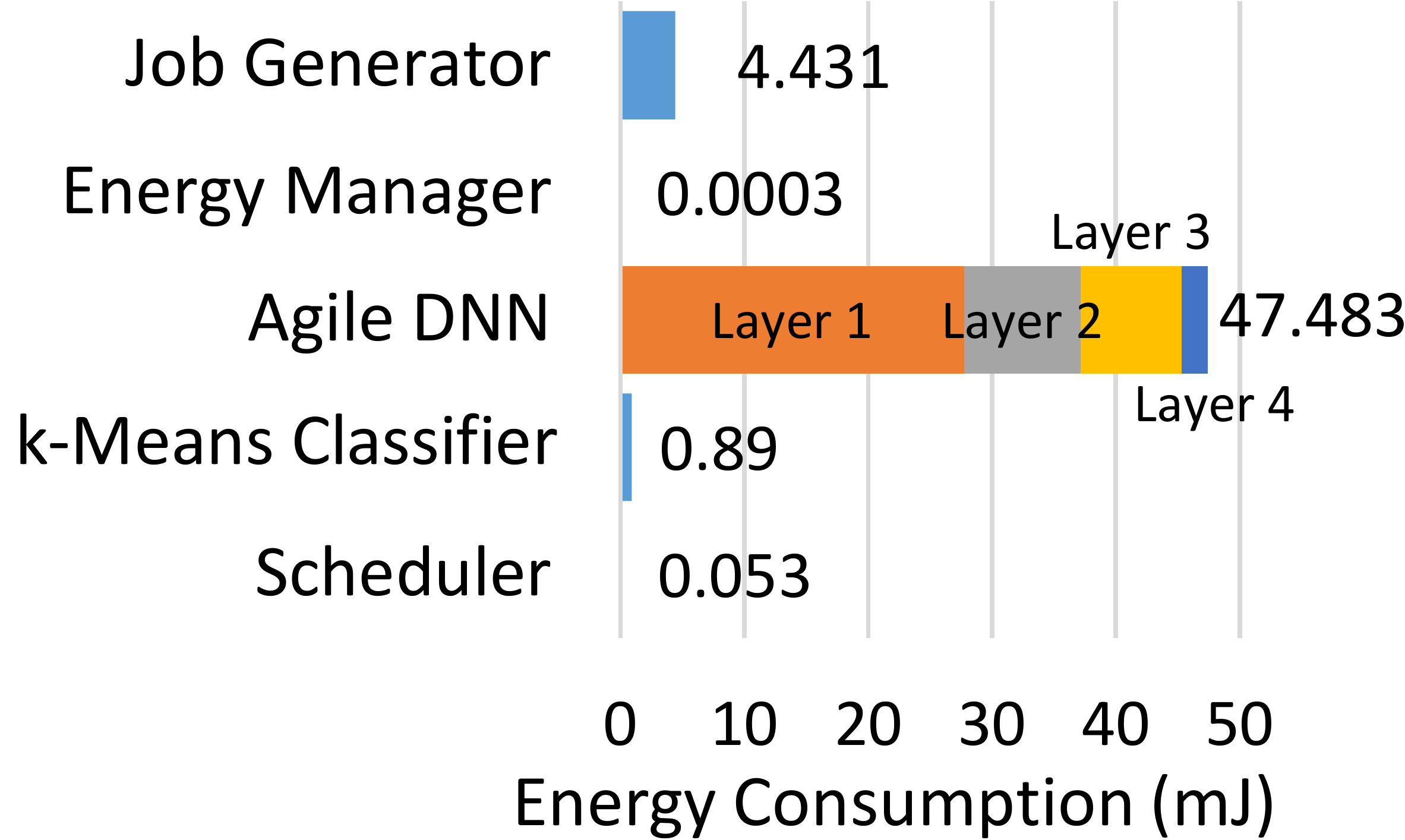}}
    \vspace{-1em}
    \caption{Overhead of \Sys.}
    \vspace{-1em}
    \label{fig:overhead}
\end{figure}

Figure~\ref{fig:overhead} shows the overhead of different components of \Sys described in Section~\ref{sec:overview}. To measure the execution time and energy consumption, we use the TI eZ-FET debug probe with EnergyTrace++~\cite{EnergyTrace}, which provides milliseconds and $\mu$J resolution data. We isolate each component of \Sys and report the average overhead from five repeated measurements.  To measure smaller overheads we repeat the experiment for multiple iterations (e.g., 2000 iterations for energy manager) and report the mean overhead of a single execution. We use a persistent power source during overhead measurements. 

The job generator reads 1s audio data from the microphone, performs FFT, and writes it to the FRAM in 1.325s. The first convolution layer (ESC-10 network of Table 1) has 2.6$\times$-3.6$\times$ higher execution time than other convolution layers due to larger input dimension. Using max-pool with stride decreases the input size, inference time, and energy consumption at each layer. The last fully-connected layer performs 50\% less multiplications than the previous layer and thus has a lower cost. Each job executes the semi-supervised $k$-means classifiers at most four times. It is 14$\times$ faster and 13$\times$ more energy-efficient than executing the whole DNN. Execution of the $k$-means classifier includes performing the utility test, classifying with k-means classifier and updating the model centroids for adaptation. For N examples in the system, the scheduler kicks in 
$4N$ times and the overhead of this specific example with three jobs are 3.72 ms and 636$\mu$J, which is less than 1\% of the overall cost of processing an example. The energy manager has negligible cost and runs once every time the scheduler executes. 




\begin{figure}[!htb]
\begin{minipage}{0.50\textwidth}
\centering
    \includegraphics[width=.9\textwidth]{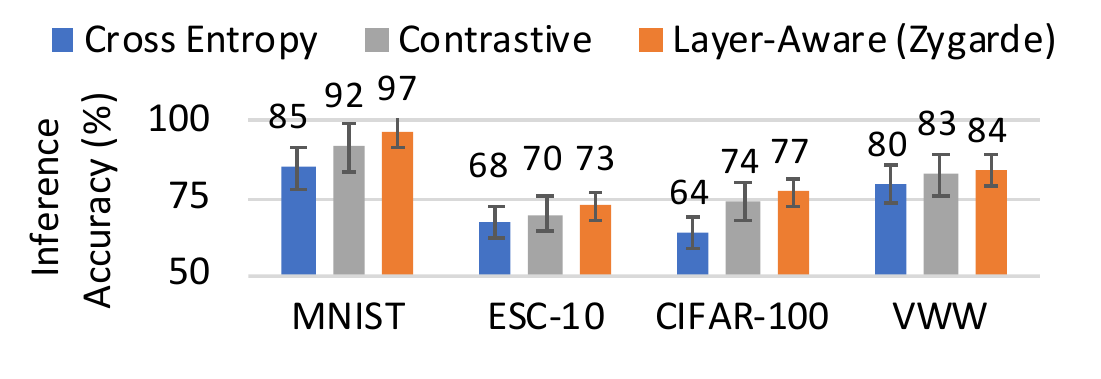}
    \includegraphics[width=.88\textwidth]{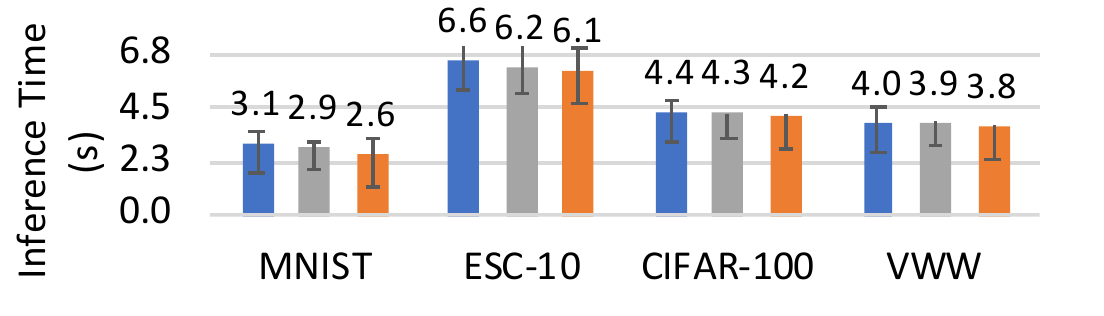}
    \vspace{-1em}
    \caption{Comparison of Loss Functions with Early Exit.}
    \label{fig:layer_aware}
\end{minipage}
\hspace{1em}
\begin{minipage}{0.46\textwidth}
\centering
    \includegraphics[width=.95\textwidth]{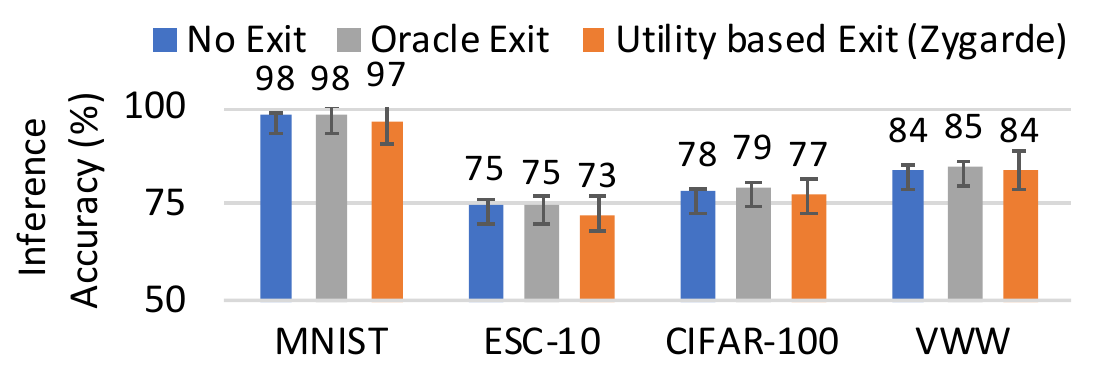}
    \includegraphics[width=.93\textwidth]{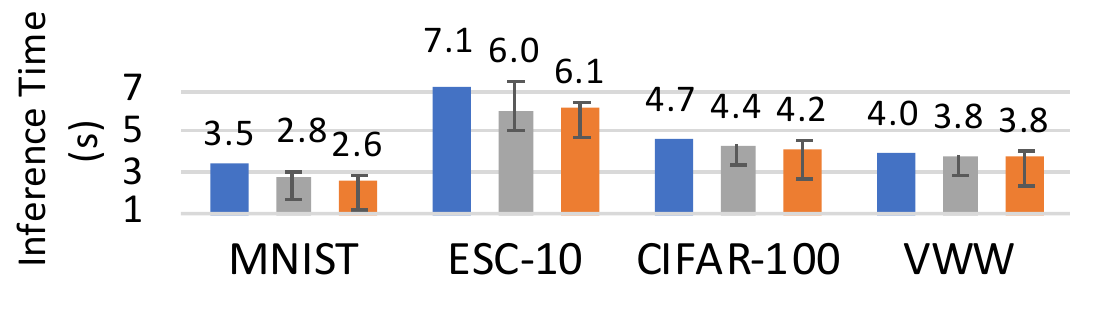}
    \vspace{-1em}
    \caption{Comparison of the Termination Policies.}
    \label{fig:utility_function}
\end{minipage}
\vspace{-1em}
\end{figure}

\subsection{Effect of Layer-Aware Loss Function}
\label{sec:layer_eval}
In Figure~\ref{fig:layer_aware}, we compare our proposed layer-aware loss function with cross-entropy loss~\cite{zhang2018generalized} and contrastive loss~\cite{islam2019soundsemantics} functions when early termination is in action. Since loss functions are equally applicable to both persistently-powered and energy-harvested systems, we conduct this experiment in a persistently-powered setting.
We train three agile DNNs with three different loss functions that have the same network structure, hyper-parameters, and training dataset. All three networks use the proposed utility test where the utility-threshold is determined during training.

Though the loss functions achieve similar accuracy ($\approx$98\% for MNIST and $\approx$75\% for ESC-10) without early termination, their performance varies when early termination is applied. Note that, the inference accuracy of ESC-10 suffers due to downsampling of the 5s and 44KHz data samples to 1s and 8KHz data samples.
In Figure~\ref{fig:layer_aware}, the layer-aware loss function demonstrates 4.13\%-13.40\% higher accuracy than cross-entropy loss by forcing the layers to learn distinguishable features~\cite{xie2015holistically}. It also decreases the average inference time by upto 13.97\%, by executing the final layer of 14\%-26\% less jobs compared to cross-entropy loss. Layer-aware loss function further achieves 2\%-5\% higher accuracy and 2\%-9\% less average inference time than the contrastive loss function. Thus, layer-aware loss function achieves higher accuracy and lower inference time than other loss functions when early termination is active.

\begin{figure}[t]
\begin{minipage}{\textwidth}
    \centering
    \includegraphics[width=\textwidth]{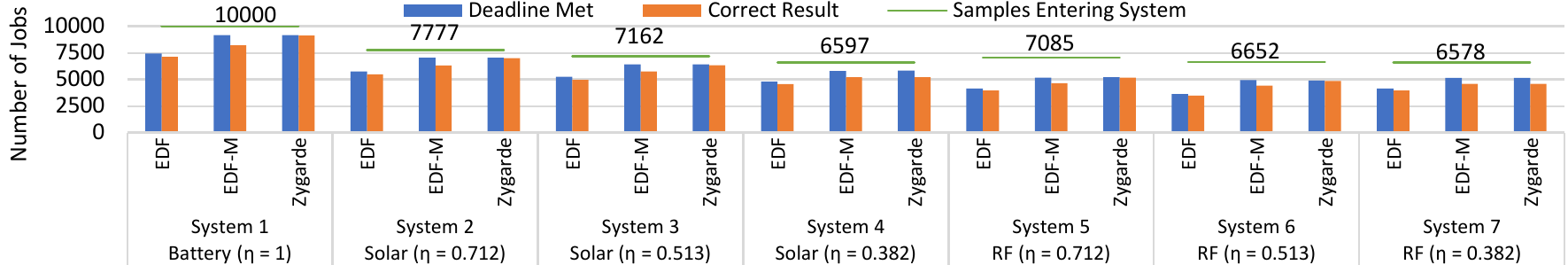}
    \vspace{-.22in}
    \caption{Real-time Scheduling for different Systems on MNIST test dataset.}
    \label{fig:schedule_mnist}
    \vspace{1em}
\end{minipage}
\begin{minipage}{\textwidth}
    \centering
    \includegraphics[width=\textwidth]{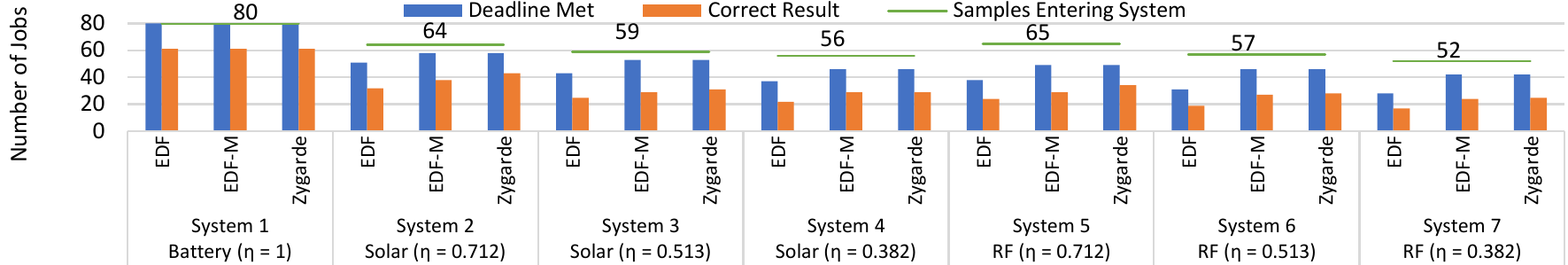}
    \vspace{-.25in}
    \caption{Real-time Scheduling for different Systems on ESC-10 test dataset.}
    \label{fig:schedule_esc}
    \vspace{1em}
\end{minipage}
\begin{minipage}{\textwidth}
    \centering
    \includegraphics[width=\textwidth]{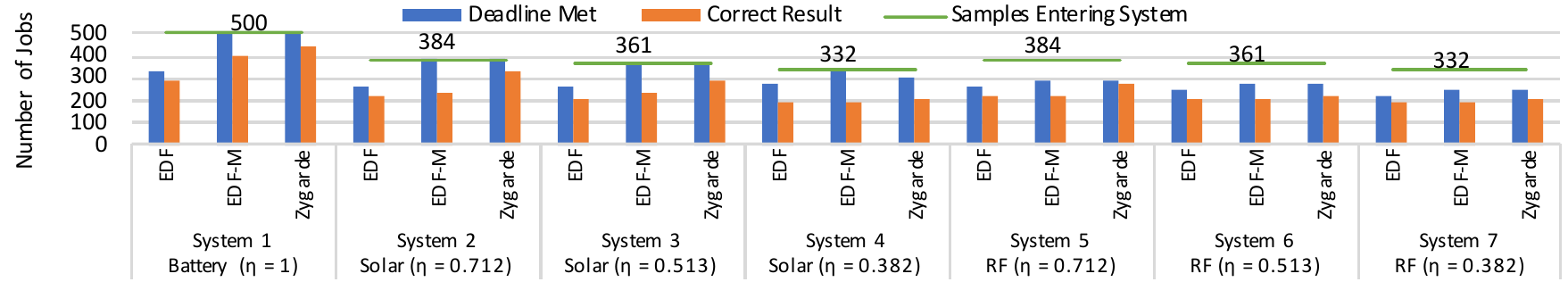}
    \vspace{-.25in}
    \caption{Real-time Scheduling for different Systems on CIFAR-100 test dataset.}
    \label{fig:schedule_cifar}
    \vspace{1em}
\end{minipage}
\begin{minipage}{\textwidth}
    \centering
    \includegraphics[width=.99\textwidth]{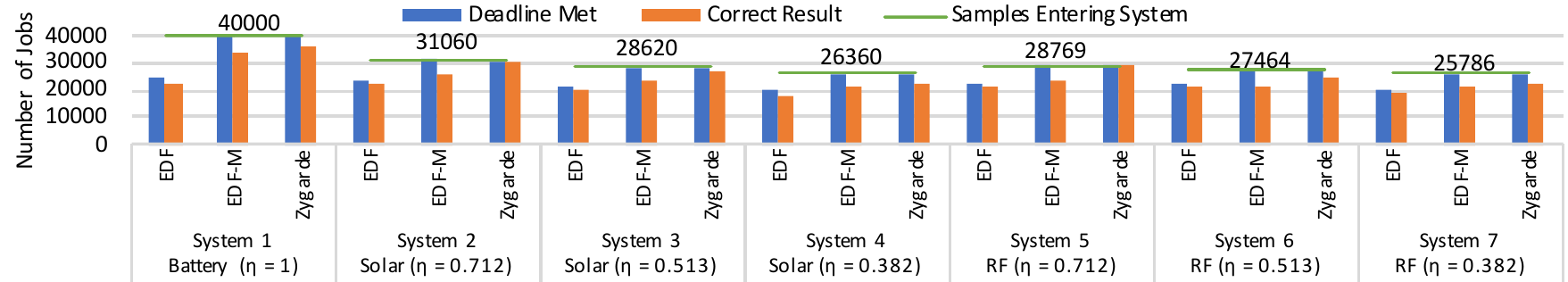}
    \vspace{-0.75em}
    \caption{Real-time Scheduling for different Systems on Visual Wake Word test dataset.}
    \label{fig:schedule_vww}
\end{minipage}
\vspace{-1em}
\end{figure}

\subsection{Effect of Early Termination}
In Figure~\ref{fig:utility_function}, we evaluate the proposed utility test by comparing it with a system that does not implement early exit and an oracle that knows the exact number of units needed for each data sample. We use the same persistently-powered system and dataset as in Section~\ref{sec:layer_eval}. All of these systems use the same trained network with the layer-aware loss function. Utility-based termination (exit) achieves similar accuracy while lowering the average inference time by 4\%-26\%. The difference in accuracy between these systems is below 2.5\%. 

\subsection{Performance of the Real-Time Scheduler}
We evaluate the proposed scheduling algorithm for dynamic imprecise tasks in both persistently and intermittently powered systems for four different $\eta$-factors and two different CPU utilization described in Table~\ref{tab:scenario}. To compare our proposed algorithm, we choose earliest deadline first (EDF) and one of its variants-- earliest deadline first mandatory (EDF-M). EDF-M schedules only the mandatory portions of the jobs. We choose EDF as a baseline because it is the optimal online scheduling algorithm for sporadic tasks. Here, both \Sys and EDF-M use the proposed utility test to partition jobs into mandatory and optional units. For the fairness in comparison, successful completion of a job's  mandatory units before deadline makes the job schedulable in all algorithms. Note that, we discard a job after its deadline to avoid domino effect~\cite{sharma2014performance}.

\parlabel{Persistently Powered System.} Figure~\ref{fig:schedule_mnist} shows the performance of proposed scheduling algorithm for MNIST dataset for $T=3s$ and $D=6s$. As the CPU utilization ($U$) is greater than one, none of the schedulers can schedule all the tasks even on persistent power. However, with early termination, EDF-M and \Sys schedule 17\% more jobs. In Figure~\ref{fig:schedule_esc}, we schedule 80 jobs from the ESC-10 dataset, where $U<1$, $T=0.36$ minutes, and $D=0.72$ minutes. Persistently powered system (System 1) can schedule all the tasks with EDF, EDF-M, and \Sys. In Figures~\ref{fig:schedule_cifar} and ~\ref{fig:schedule_vww}, we schedule 500 and 40,000 jobs for CIFAR-100 and visual wake words (VWW) datasets, respectively, where the deadline is twice the period. In both cases, EDF-M and \Sys schedules all the jobs while EDF fails to do so. As successfully scheduling only the mandatory units of a job before deadline is sufficient to be schedulable, EDF-M schedules similar number of jobs as \Sys. However, \Sys achieves higher accuracy by opportunistically executing optional units.

\parlabel{Intermittently Powered Systems.}
For intermittent systems (Systems 2-7), EDF-M schedules 14.98\%-19.51\% more jobs for MNIST, 9.44\%--20.70\% more jobs for ESC, 8.59\%--33.59\% more jobs for CIFAR, and 16.97\%--24.53\% more jobs for VWW than EDF. If the utility tests were optimal, EDF-M would have produced correct results for all the scheduled jobs. However, due to the limitation of utility tests, \Sys increases the number of scheduled jobs that produce the correct results by up to 27.60\% by executing some of the optional units. We observe that, \Sys increases the performance (i.e., the number of scheduled jobs that produce correct results) from EDF-M when $\eta$ is high. With low $\eta$, the performance of \Sys and EDF-M becomes similar as no optional units are executed. It is interesting to notice that despite having the same $\eta$, solar powered systems schedule 9\% - 31\% more jobs than RF powered systems due to more available power.

\begin{figure}[h]
\vspace{-1.5em}
\begin{minipage}{0.35\textwidth}
    \centering
    \includegraphics[width=\textwidth]{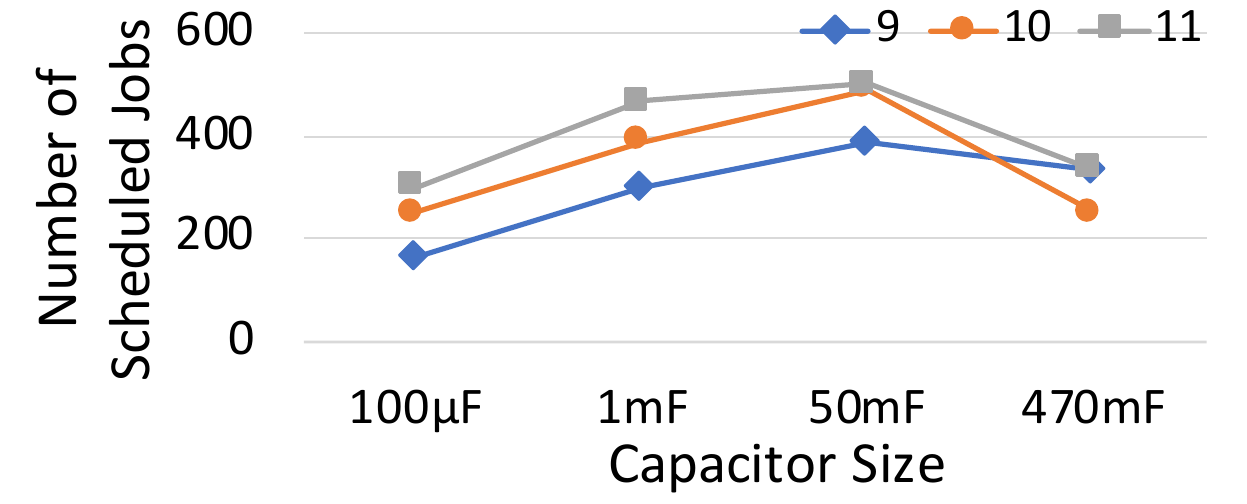}
    \vspace{-2.2em}
    \caption{Effect of Capacitor}
    \label{fig:cap}
\end{minipage}
\begin{minipage}{0.58\textwidth}
\captionsetup{type=table}
    \vspace{1em}
    \centering
    \caption{Effect of Cascaded Hierarchical Remanence Timekeeper}
    \vspace{-10pt}
    \begin{footnotesize}
    \begin{tabular}{|c|P{1.5cm}|P{1.5cm}|P{1.5cm}|P{1.5cm}|}
    \hline
    \textbf{System} & \textbf{Number of Reboots} & \textbf{Power On Time} & \textbf{Scheduled Tasks using RTC} & \textbf{Scheduled Tasks with CHRT}\\ \hline
    2 & 67 & 77.67\% & 29989 & 29980 \\ 
    3 & 1252 & 71.48\% & 27401 & 27390 \\ 
    4 & 1820 & 65.83\% & 24921 & 24897 \\ \hline
    \end{tabular}
    \vspace{0.75em}
    \label{tab:timer}
    \end{footnotesize}
\end{minipage}
\vspace{-1.5em}
\end{figure}

\subsection{Effect of Capacitor Size}

The goal of this experiment is to quantify the effect of the capacitor's size on scheduling. We use the CIFAR-100 dataset and its corresponding DNN (Table~\ref{tab:network}), and power the system from an intermittent RF energy source ($\eta$ = 0.51) at around 0.5m distance. The period of the tasks are varied between 9s to 11s and the deadline is set to twice the period. We use four different capacitors: 0.1mF, 1mF, 50mF, and 470mF. This setup and workload stress tests the system and forces the scheduler to miss the deadline when the capacitor values are too small or too high. Figure~\ref{fig:cap} shows that when the capacitor value is below 50mF, more tasks miss their deadlines as they re-execute an atomic fragment when the power goes off before its completion. On the other hand, when the capacitor value is high (e.g., 470mF), tasks miss deadline due to the extra time required to charge such a large capacitor. Hence, we choose to use a 50mF capacitor for the rest of the experiments. Note that although we empirically determine a suitable capacitor for our experiments, one can roughly estimate the optimal value of the system capacitor, $C$ by using a capacitor's energy equation, when the average input power, $P$, voltage across the capacitor, $V$, and the difference between the deadline and the total execution time of the task, $\delta T$, is known: $C = \sqrt{\frac{2 P \delta T}{V^2}}$.

\subsection{Effect of Remanence Clock}

Keeping track of the time is crucial for a real-time scheduler and it is a hard problem, in general, for  batteryless systems. To keep track of time reliably across power failures, recently, a batteryless remanence clock, namely the \emph{Cascaded Hierarchical Remanence Timekeeper} (CHRT)~\cite{de2020reliable} has been proposed for intermittently-powered systems. The CHRT clock has three modes or tiers. Its tier-1 yields near-perfect time-keeping accuracy, but has a range of only 100ms. On the other hand, the tier-3 offers 1s resolution, 100s range, and reports accurate time 80\% of the cases, while reporting +1s error for the rest of the time and rarely shows +2s, -1s or -2s error. We implement this clock following the open source hardware design (see Figure~\ref{fig:hardware}) and use it to power \Sys -- to implement a completely batteryless system. We evaluate the effect of batteryless CHRT clock on \Sys's scheduler and compare it to the performance of \Sys when it uses battery-powered RTC.

Table~\ref{tab:timer} shows the number of tasks meeting deadlines for both types of clocks for the systems 2--4 (see Table~\ref{tab:scenario} for definitions). We do not show results for systems 5--7, which are powered by RF harvesters and require using CHRT tier-1 (which is optimized for RF), since the results are identical for both CHRT and RTC. We observe that the number of missed jobs increases with the number of reboots due to intermittent energy. Upon investigating the cause we find that during the positive error of the CHRT clock, the scheduler either reports the missed deadlines or terminates a job early, as it mistakenly thinks that the deadline has passed, and thus, continuing to execute these tasks is a waste of time. During negative error of the CHRT clock, the scheduler schedules a job despite the fact that it missed the actual deadline and triggers a domino effect that results in more tasks missing their deadlines. However, CHRT shows negative error $< 3\%$ time and often it compensates for a positive error. Overall, the loss of schedulable tasks due to the use of a batteryless clock is below $0.1\%$.




\section{Real-World Application Evaluation}
\label{sec:real_evaluation}

In the previous section, we compared the performance of \Sys with different baseline algorithms. In this section, we observe \Sys in two real-world applications. In the first application, we perform acoustic sensing and show how different scenarios affect the system performance. In the second application, we compare the performance of \Sys with a state-of-the-art intermittent DNN inference system~\cite{gobieski2019intelligence} for visual sensing tasks in a real-world setting.

\subsection{Acoustic Sensing}
 \label{sec:real}

\parlabel{Experimental Setup.} In this section, we evaluate \Sys in real-world uncontrolled experiments using six audio event detection applications. Due to the presence of background noise and multiple audio events in the data, these applications require DNN-based features for audio event representation and classification. Existing works show that DNN performs significantly better than threshold or classic machine learning-based audio event detectors in real-life noisy environments~\cite{kons2013audio, esccompare}.

Table~\ref{tab:real_setup} shows the the application environment, energy source, harvester placement, cause of energy intermittence, target event and other events present in the environment for the six applications. 
Each applications runs for 10 minutes and the audio sensor samples every two seconds. We play recorded sound, that are not used during training, 10 times from a speaker as the positive example. The relative deadline of the jobs are 3s which is the required execution time for the whole model. The agile DNN, consisting of a convolution layer and two fully connected layers, has an execution time that varies between 1.7s and 3s, depending on early termination. As it is not possible to ensure that each audio event of the target classes falls neatly into one second buckets, we combine the outputs of two consecutive jobs by taking their logical OR.   

\begin{table*}[t]
\centering
\caption{Real-life evaluation setup.}
\vspace{-10pt}
\resizebox{\textwidth}{!}{%
\begin{small}
\begin{tabular}{|l|l|l|l|l|l|l|}
\hline
\textbf{Application}  & \textbf{Energy Source} & \textbf{Harvester Placement} & \textbf{Cause of Intermittence}   & \textbf{Target Event} & \textbf{Other Events}          \\ \hline

Car Detector          & Solar         (74Klx to 111Klx)    & Pavement    & Vehicle on the closest lane & Car Honk     & Silence, Dog, Human Voice, Car            \\

Barking Dog    & Solar (2Klx to 18Klx)    & Under the Tree          & People, objects and cloud   & Dog Bark     & Silence, Car, Car Honk,  Voice       \\

People Detector & Solar         (1Klx to 5 Klx)    & Edge of the Railing & People and cloud            & Voice  & Silence, Car, Honk, Dog Bark          \\

Baby Monitor               & RF (-0.48dB to -1.66dB)    & On the Desk             & Change of Distance                  & Crying Baby   & Silence, Voice, Washer, Printer     \\

Laundry Monitor             & RF (-0.48dB to -1.91dB)      & On the Counter          & Change of Distance                  & Washer Status       & Silence, Voice, Cry, Printer \\

Printer Monitor               & RF (-1.59dB to -1.91dB) & On the Desk             & Change of Distance                  & Printer Status     & Silence, Voice, Crying Baby, Printer \\ 
\hline
\end{tabular}
\end{small}
}
\label{tab:real_setup}
\vspace{-2em}
\end{table*}

\begin{figure*}[t]
    \centering
    \includegraphics[width=\textwidth]{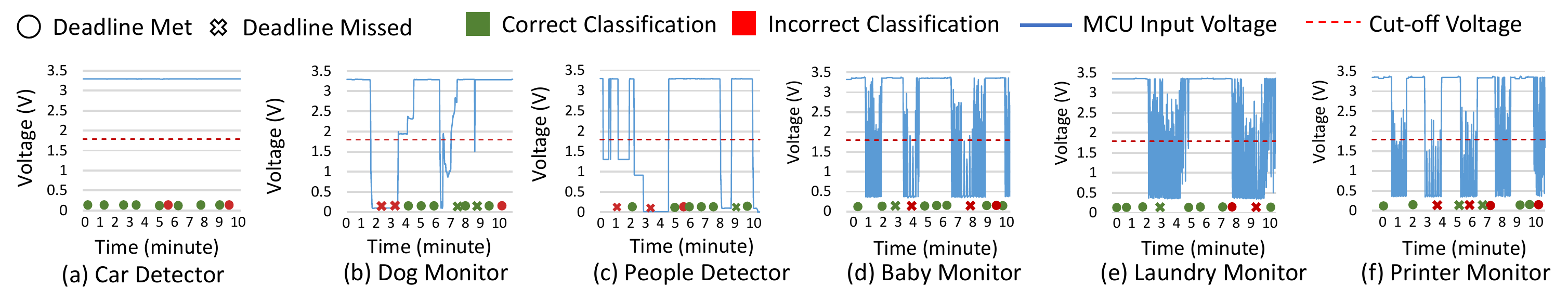}
    \vspace{-2em}
    \caption{Real-life evaluation of \Sys for acoustic event detection.}
    \label{fig:real}
    \vspace{-1.5em}
\end{figure*}
We use two energy harvesters: solar and RF. The solar energy harvester is affected by outdoor influences such as passing vehicles. We vary the distance between the RF transmitter and the receiver to test the applications under different levels of noise and interference.





\parlabel{Results.} In Figures~\ref{fig:real}(a)-(f) we plot the MCU's input voltage, the cut-off voltage, the classifier's output, and deadline misses for the six applications over time. Findings from this experiments are as follows.  



The car detector in Figure~\ref{fig:real}(a) always harvests sufficient energy from the sun and meets the deadline for all jobs. However, it misclassifies twice when both pedestrians (talking) and cars are in the scene due to the limitations of the classifier. 
The dog monitor in Figure~\ref{fig:real}(b) experience intermittency due to people blocking the sun. It misses two target events due to the lack of sufficient energy to read the sensor data and misclassifies one event due to the limitation of the classifier. For two audio events, the applications experience deadline misses despite doing accurate classification because of the limitation of the utility test.
For similar reasons, the people detector in Figure~\ref{fig:real}(c) fails to sense two events and misclassifies one. 

The baby monitor in Figure~\ref{fig:real}(d), powered with an RF harvester, does not harvest enough energy to read audio samples during one audio event. It also fails to finish execution of mandatory units within the deadline for one event. Due to the limitation of the utility test, it misclassifies one event and misses deadline of another. The Laundry monitor in Figure~\ref{fig:real}(e) misclassifies one event and misses the deadline for two. The printer monitor in Figure ~\ref{fig:real}(f) experiences the highest intermittence, misses four deadlines and misclassifies three events.

A number of interesting observations from these experiments are: 
(1) a shorter power-off period decreases the number of event misses, e.g., the solar powered dog monitor misses more events than the laundry monitor despite having less frequent reboots due to insufficient power supply;
(2) a shorter continuous energy results in more deadline misses, as evident in dog monitor and printer monitor applications;
(3) deadline and target event misses depend on the harvested energy and the accuracy of the utility test, whereas the classification accuracy relies on the competence of the classifier and the accuracy of the utility test, e.g., the car detector misclassifies due to the limitation of the classifier, whereas the dog monitor misses the deadline of two correctly classified samples due to the inaccuracy of the utility test.

\subsection{Visual Sensing}
\parlabel{Experimental Setup.} We evaluate the performance of \Sys in a multi-tasking scenario having two visual recognition tasks: traffic sign recognition and shape recognition. Both DNNs have two convolution layers and two fully-connected layers, but the convolution layers of the sign recognizer has 8 and 16 filters, whereas the convolution layers of the shape recognizer has 4 and 8 filters. The shape recognizer's execution time is about half of sign recognizer's execution time, and hence, it has a smaller relative deadline. After capturing an image, a sign detection job is created and inserted into the job queue, followed by a shape detection job.

\begin{figure*}[!htb]
    \centering
    \includegraphics[height=1in]{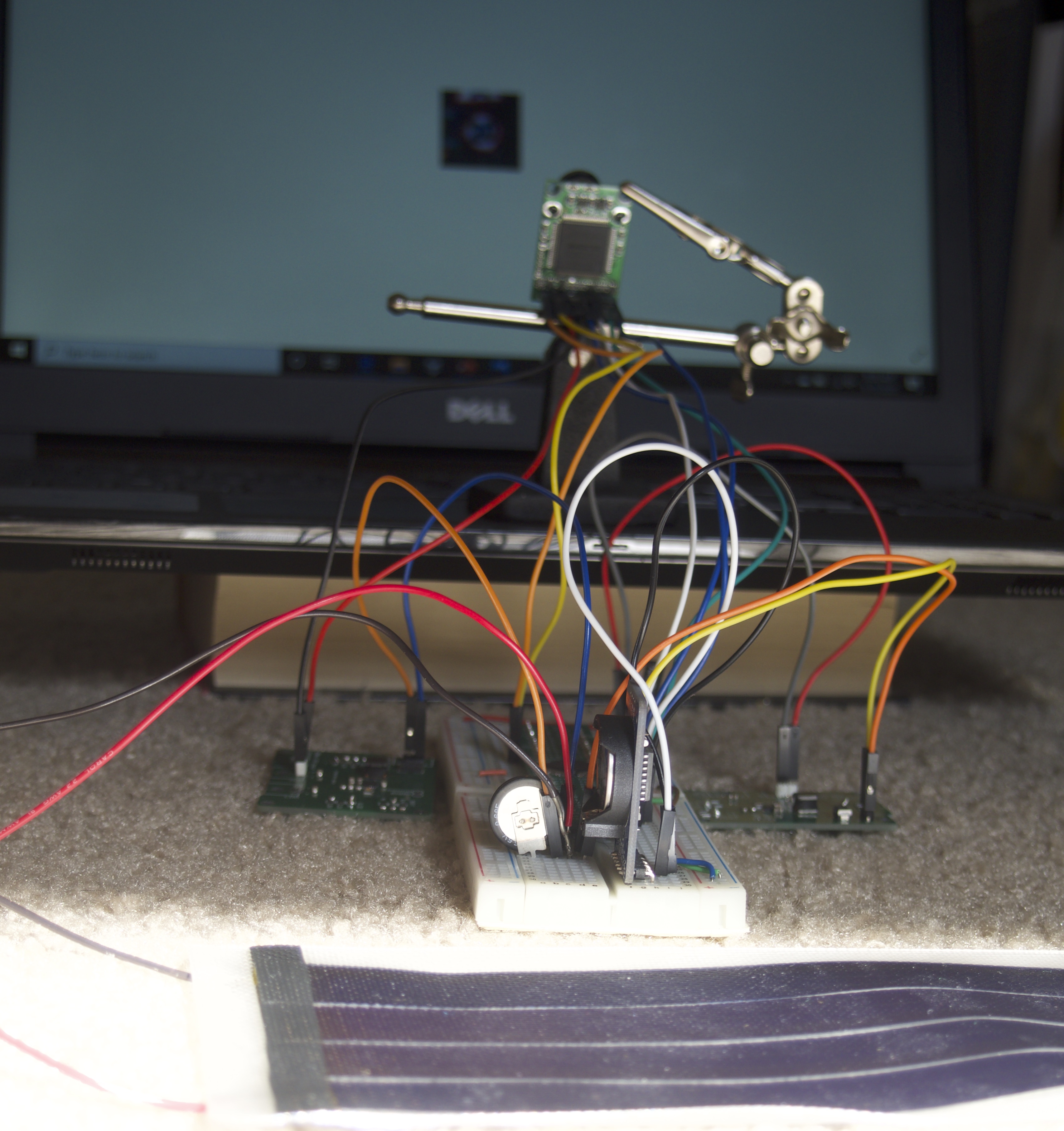}
    \hspace{2em}
    \includegraphics[height=1in]{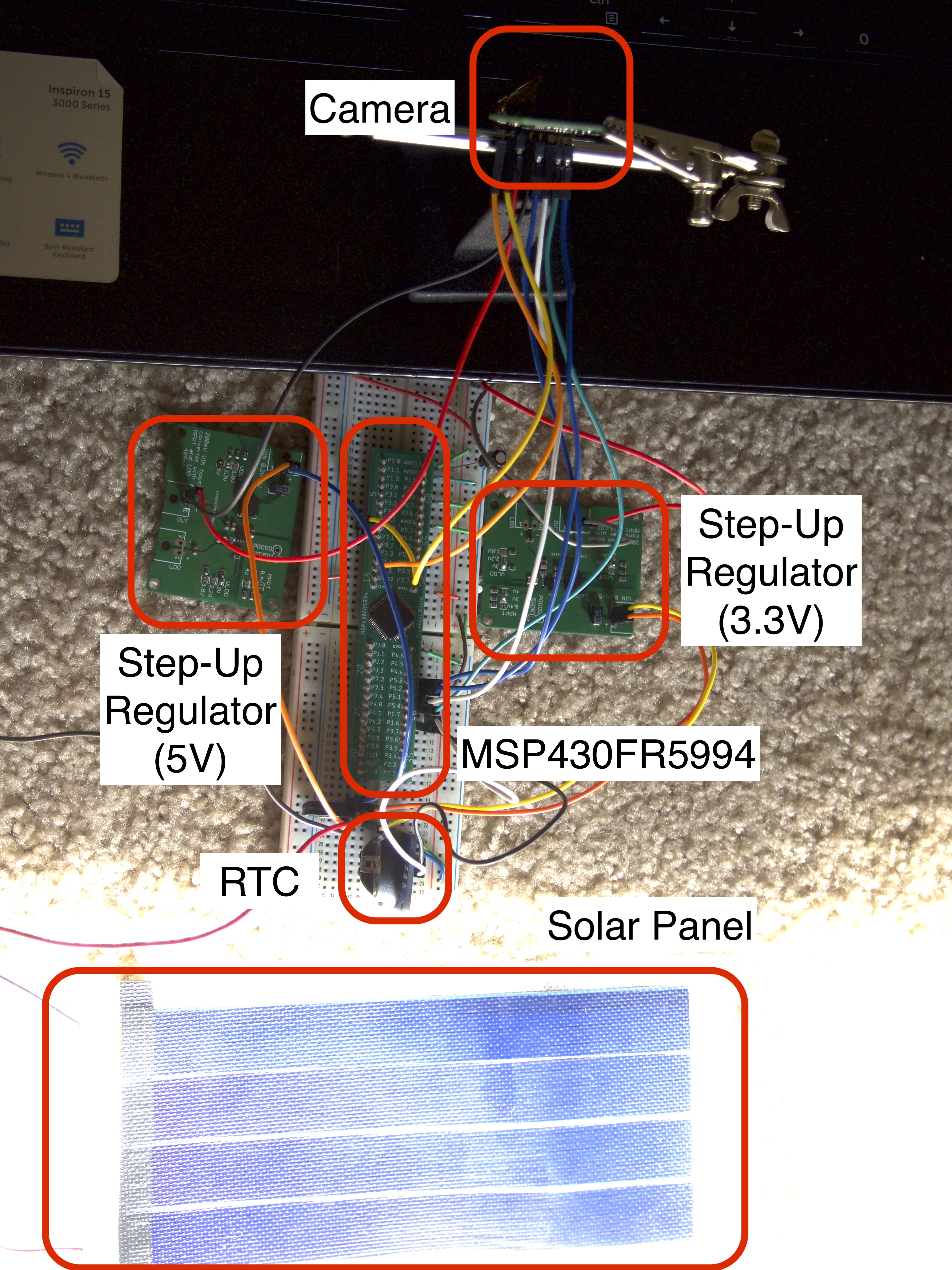}
    \hspace{4em}
    \includegraphics[height=1in]{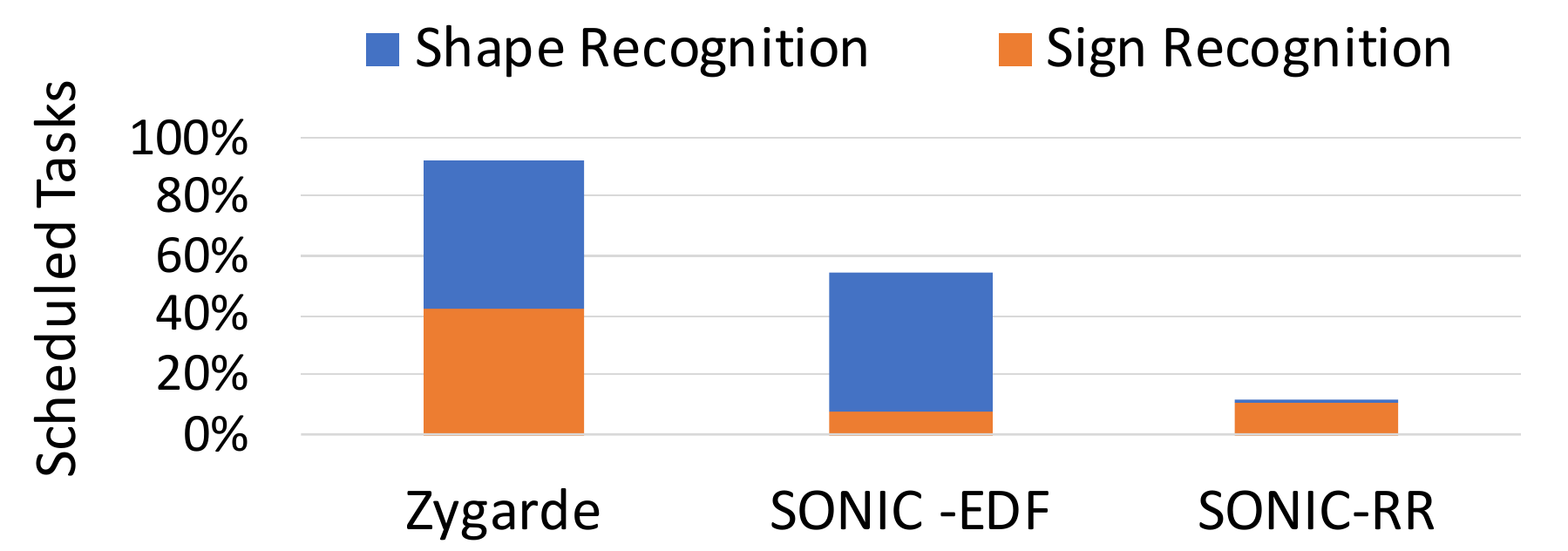}
    \caption{(a) Experimental setup. (b)  Percentage of captured events that meet the deadline.}
    \label{fig:real_camera}
\end{figure*}

Figure~\ref{fig:real_camera}(a) shows the setup for this experiment. We use a 2MP OV2640 camera sensor and capture the test images from the GTSB~\cite{Stallkamp2012} dataset displayed on the screen of a laptop. We use 80\% of the dataset for training and the remaining 20\% for testing. We annotate the dataset to label the shape of the sign. We use a solar energy harvester to power the system and acquire 5V and 3V power lines for the camera and the MSP430, respectively, by using two voltage regulators. The camera module requires 4s to capture an event but works in parallel with the MSP430 which uses DMA.

\parlabel{Results.}
In Figure~\ref{fig:real_camera}(b), we compare the performance of \Sys against SONIC's~\cite{gobieski2019intelligence}, which does not implement early termination and uses either an EDF or a round-robin (RR) scheduler. We observe that due to the high energy demand of the camera,  37\% of the events are missed and do not enter any of the three systems. Although SONIC-EDF schedules 55\% of the jobs that enter the system, it is partial towards the shape recognition jobs since they have earlier deadlines. By choosing the sign recognition job, which has higher execution time, SONIC-RR does not spare sufficient time to execute shape recognition job. SONIC-RR schedules only 11\% jobs that enter the system in total, among which, only 1\% are shape recognition jobs due to the shorter relative deadline of the shape recognition task. By performing imprecise computing with early termination, \Sys assigns different priority to the same job at different units. Thus, \Sys switches between jobs from different tasks and enables fairness. \Sys schedules 93\% of the jobs that enter the system, where 43\% are sign recognition jobs and 50\% are shape recognition jobs. \Sys achieves 61\% and 85\% classification accuracy for sign and shape recognition, respectively, which is is within 2\% of the baselines' that execute the DNNs end-to-end.

\section{Related Work}
\label{sec:related}

This section describes the existing literature on intermittent computing, timeliness in batteryless systems, DNN compression and partial execution (Anytime Neural Networks), energy harvesting system model, and real-time scheduling for imprecise and mixed critical task set. 

\parlabel{Intermittent Computing.}
Intermittently powered systems experience frequent power failures that reset the software execution and result in repeated execution of the same code and inconsistency in non-volatile memory. Previous works address the progress and memory consistency using software check-pointing~\cite{ransford2012mementos, maeng2018adaptive, hicks2017clank, lucia2015simpler, van2016intermittent, jayakumar2014quickrecall, mirhoseini2013automated, bhatti2016efficient}, hardware interruption~\cite{balsamo2015hibernus, balsamo2016hibernus++, mirhoseini2013idetic}, atomic task-based model~\cite{maeng2017alpaca, colin2016chain, colin2018termination} and, non-volatile processors(NVP)~\cite{ma2017incidental, ma2015architecture}. Recently SONIC~\cite{gobieskiintermittent, gobieski2019intelligence} proposes a unique software system for intermittent execution of deep neural inference combining atomic task-based model with loop continuation. Despite being used for intermittent execution by \Sys, SONIC does not handle time-aware, partial execution of DNN with model adaptation like \Sys.

\parlabel{Timeliness of Batteryless Systems.}
Prior works on intermittent computing propose runtime systems to increase the likelihood of task completion by finding optimum voltage~\cite{buettner2011dewdrop}, adapting execution rate~\cite{sorber2007eon, dudani2002energy}, and discarding stale data~\cite{hester2017timely}. However, none of these works consider the accuracy/utility of the target application or deadline-aware execution of tasks. Some works in wireless sensors~\cite{zhu2012deos, moser2006lazy, moser2007real} have addressed scheduling, but none of them consider the higher computation load of inference tasks. INK~\cite{yildirim2018ink} proposes a reactive kernel that enables energy-aware dynamic execution of multiple threads. Note that INK does not consider DNN tasks or utilize partial execution of tasks for increasing schedulability. Unlike INK, which schedules kernel threads and only one data sample at a time, \Sys~ schedules multiple data samples present in the job queue. Recent work on real-time in intermittent systems~\cite{islam2020scheduling} explores this problem for only periodic tasks and unlike \Sys~ does not considers the impreciseness of the taskset. Other works focus on maintaining time through power loss~\cite{de2020reliable, rahmati2012tardis, hester2016persistent} by exploiting discharge rate of SRAM and capacitors. Our work is complementary to these works and relies on these techniques for timekeeping.  

\parlabel{Compression and Partial execution of DNN.}
Recent works reduce the cost of DNN inference by pruning and splitting models~\cite{han2015deep,nabhan1994toward, yao2017deepiot, yao2018rdeepsense, yao2018apdeepsense, yao2018fastdeepiot}, reduction of floating-point and weight precision~\cite{johnson2018rethinking, de2017understanding, han2016eie}, and factorization of computation~\cite{bhattacharya2016sparsification, chollet2017xception, nakkiran2015compressing,szegedy2017inception, szegedy2015going, szegedy2016rethinking, wang2017factorized}. Though these works are crucial for enabling fast DNN execution, they alone are not sufficient for batteryless systems. Binary networks~\cite{courbariaux2015binaryconnect, hubara2016binarized,rastegari2016xnor, courbariaux2017binarynet} are not suitable for batteryless systems due to the higher number of required parameters~\cite{gobieski2019intelligence}. 

Recent works propose early exit during DNN inference~\cite{teerapittayanon2016branchynet, bolukbasi2017adaptive, figurnov2017spatially, leroux2017cascading}. However, they are not sufficient for highly constrained batteryless systems due to the large termination overhead. In most cases, the terminations require execution of a neural layer requiring $45\times$ more execution cycles than performing utility test and classification with \Sys~. Moreover, these approaches lack model adaptation capability which is required for life-long sensing. Some works on anytime neural networks depends on module selection ~\cite{teja2018hydranets, odena2017changing, huang2017multi}, dynamic layer pruning during inference~\cite{veit2018convolutional, huang2016deep, bolukbasi2017adaptive, harvey2002skipnet, huang2017multi} and depth and width adjusting techniques~\cite{yu2019universally, karayev2012timely, karayev2014anytime, hu2019learning, kim2018nestednet} for anytime prediction. However, none of these works have considered the the effect of energy intermittence and they are yet to be modeled as imprecise task. 

\parlabel{Modeling Energy Harvesting Systems.}
Previous works on energy harvesting (EH) modeling of a specific energy source have achieved promising results in predicting available energy~\cite{crovetto2014modeling, sharma2018modeling, jia2018modeling}. Some other works have focused on analytically model the trade-off associated with length of history to maximize forward propagation~\cite{san2018eh}. However, none of the prior works are generalizable and fails to model energy harvesting systems irrespective of energy source. In \Sys, we provide a single metric, $\eta$ factor that models the predictability of an energy harvesting system irrespective of the source. 

\parlabel{Real-time Scheduling for Imprecise and Mixed Critical Task Set.}
Previous works on imprecise computing only considered the task models where mandatory-optional partitions are fixed and known a priori~\cite{liu1991algorithms, shi2009hash}. Existing works on Quality of Service (QoS) based resource management~\cite{rajkumar1997resource, lee1999scalable} do not handle the data-dependant dynamic relationship between quality and required unit for each job in ~\Sys. Though, ~\Sys can also be formulated as a mixed-critical system, the changing utility of jobs at runtime and intermittent energy makes it more suitable as a imprecise-computing system. Previous works on mixed-critical systems~\cite{baruah2011response, baruah2010towards, vestal2007preemptive, li2010algorithm, dorin2010schedulability, buttazzo1995value} propose scheduling schemes for predetermines critical levels, which is a characteristic of a task. On the other hand, utility in ~\Sys varies for each job.


\section{Discussion}
\label{sec:discussion}

\subsection{Importance of DNNs}

For batteryless sensing systems, the inference accuracy dictates the response time and the energy-efficiency of the system~\cite{gobieski2019intelligence}. Due to very high energy cost of wireless communication, these systems have to implement a large capacitor -- which takes several minutes to charge -- in order to send just one data packet. Hence, every false positive wastes significant amount of energy and time. This is why, DNNs are preferred over less accurate traditional classifiers such as Support Vector Machines (SVM), K-Nearest Neighbours (KNN), k-means, and Random Forest. Table~\ref{tab:model_acc} shows that DNNs are 1\%--15\% more accurate than the traditional classifiers. 



\begin{table}[!htb]
\begin{small}
\caption{Classification Accuracy for Different Models.}
\vspace{-10pt}
\begin{tabular}{|l|c|c|c|c|}
\hline
\textbf{Classifier}  & \textbf{MNIST} & \textbf{ESC-10} & \textbf{CIFAR-100} & \textbf{VWW}\\ \hline
KNN                             & 92\%~\cite{ebrahimzadeh2014efficient}        & 40\%   & 55\%     & 60\%   \\ 
K-means                         & 93\%~\cite{ebrahimzadeh2014efficient}     & 41\%        & 50\% & 59\% \\ 
Random Forest                             & 93\%~\cite{ebrahimzadeh2014efficient}           & 25\%          & 29\% & 62\% \\ 
SVM                             & 96\%~\cite{ebrahimzadeh2014efficient}           & 50\%          & 51\% & 69\% \\ 
CNN (No Early Termination) & 98\%        & 75\%     &78\% &84\%  \\ 
CNN (Early Termination)    & 97\%        & 73\%  &77\%   &84\%    \\ \hline
\end{tabular}
\end{small}
\label{tab:model_acc}
\end{table}

\subsection{Generic Utility Functions}

In \Sys, the utility function provides an estimate of how confident the cluster-based classifier is. For a different type of classifier, although the proposed utility function may not be directly applicable, the general principle behind the utility function remains the same. For some classifiers~\cite{alpaydin2020introduction}, e.g., support vector machine and K nearest neighbour, the distance of the input data point from the decision boundary or the neighbours can be used to design the utility function that is similar to \Sys's. For classifiers that provides a probability distribution over all classes as the output~\cite{alpaydin2020introduction}, e.g., neural networks, na\"ive bayes, and logistic regression, we recommend using the entropy~\cite{cheng1999entropy} of this distribution as the utility function, i.e., $U = -\sum_{i=1}^c p_i {\log}_2 p_i$, where $p_i$ is the probability of the input being in class $i$ and $c$ is the total number of classes. A higher entropy indicates that the probability of the input belonging to some class is higher than the rest of the classes, whereas a lower entropy indicates similar probability across all classes.


\begin{figure}[!htb]
\begin{minipage}{0.42\textwidth}
\centering
\includegraphics[height=1in]{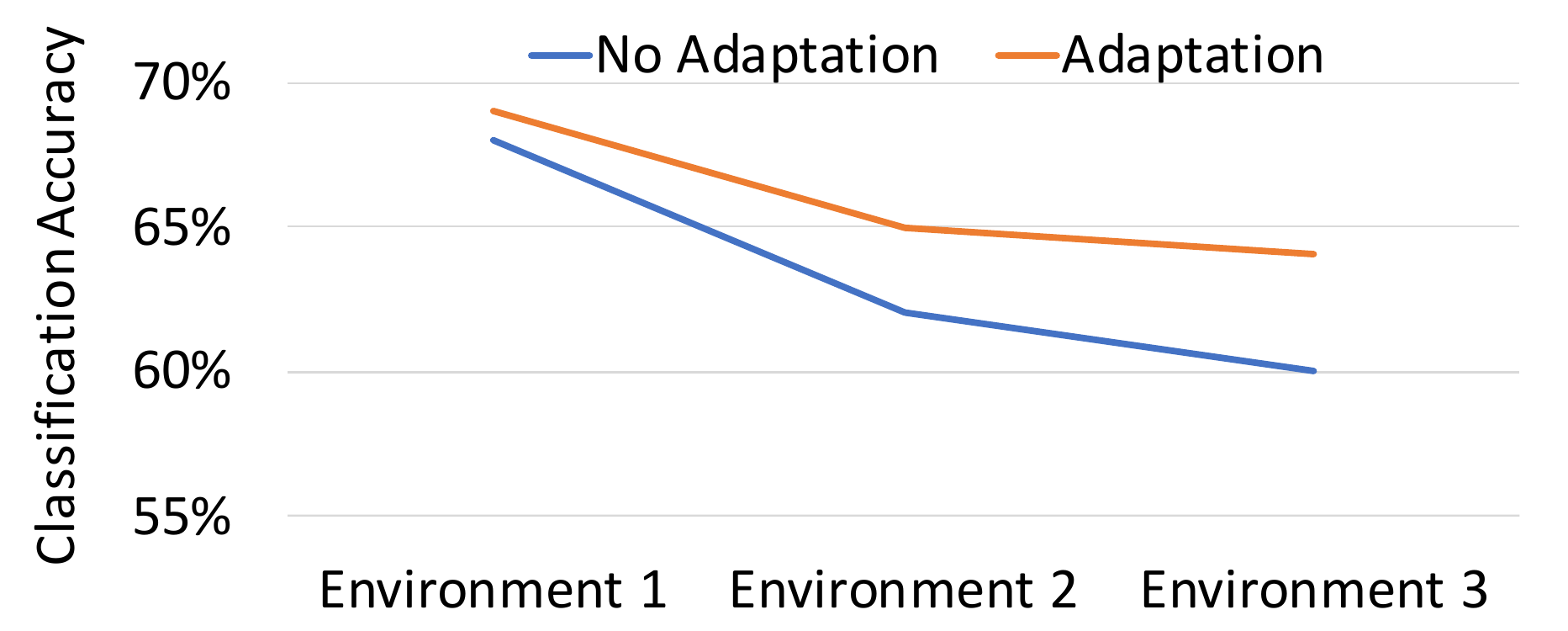}
    \caption{Performance gain due to adaptation.}
\label{fig:adaptation}
\end{minipage}
\hspace{1em}
\begin{minipage}{0.4\textwidth}
\centering
\includegraphics[height=1.1in]{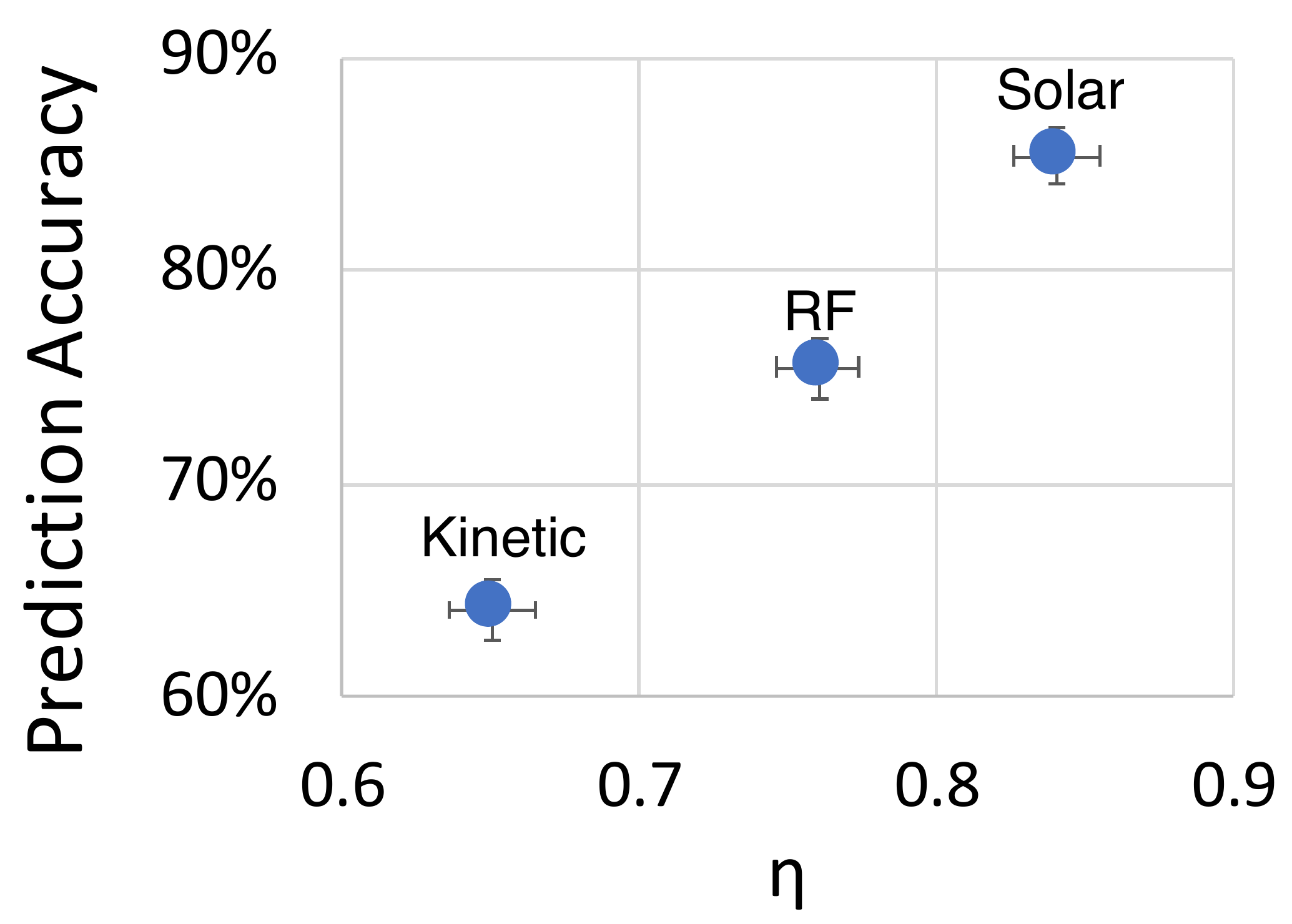}
\vspace{-1em}
\caption{Validation of $\eta$-Factor.}
\label{fig:eta_val}
\end{minipage}
\vspace{-1.5em}
\end{figure}

\subsection{Adapting the $k$-Means Classifiers}

In \Sys, we adapt the cluster-based classifiers at runtime since a classifier running on a perpetually-powered system is likely to encounter shifts in the input data distribution over its extended lifetime. To enable this, we implement a simple strategy where the cluster centroids are updated by taking the weighted average of the current centroid and the new data point. By assigning more weights to the current centroid, we ensure that the adaptation process is gradual, and is not affected by a few outliers. This strategy has both benefits and limitations.        

The benefit of cluster adaptation is that if a system is trained and tested in different environments, unless measures are taken to adapt the classifier, its accuracy drops. We conduct an experiment to quantify this. 
We first divide the ESC-10 audio dataset in to 80\% training and 20\% testing subsets. Then we record only the testing subset in three different environment, i.e., lab, hall, and office. The training subset, 80\% data, is recorded only in environment 1 and is used to train the agile DNN and the initial k-means classifier. We test the accuracy of the classifier on the testing subset from environment 1 (lab), followed by testing the accuracy on the testing subset from environment 2 (hall), followed by testing the accuracy on the testing subset from environment 3 (office).
We repeat this experiment with and without the cluster adaptation step of \Sys. Figure~\ref{fig:adaptation} shows the result. Without the adaptation, \Sys loses 8\% accuracy due to the environment changes. More than half of this lost accuracy is gained back when \Sys enabled cluster adaptation.


Two major limitations of this approach are: (1) the adaptation process being slow, if the environment changes rapidly, the system may not be able to adapt fast enough. By adjusting the weights assigned to the new data, this problem can be addressed; (2) the proposed adaptation process is robust to only a certain types of distribution shifts, e.g., translation and rotation of feature spaces, where the relative distances of the cluster heads do not change significantly. However, if the shift in the data distribution in the new environment is complex and/or non-linear, this simple threshold-based cluster adaptation approach may not work. To deal with this, a more sophisticated approach that normalizes the effect of domain shifts in data~\cite{hoffman2017cycada, mathur2019mic2mic} has to be employed by adding an extra layer of computation prior to the clustering step.   




\subsection{Limitation of $\eta$ Factor}

In \Sys, we estimate the $\eta$-factor offline, using energy harvesting traces of the systems that we experiment with. The modeling accuracy of $\eta$ largely depends on the length of this empirical study -- which must be long enough to capture the variability in harvested energy. Note that this variability depends not only on the energy source, but also on how the system is used by a user or how/where it is deployed. Hence, prior knowledge about the system's energy usage pattern and/or the experience of the system designer are crucial to determining a reasonable study duration to obtain an accurate estimate of the $\eta$-factor.           

A limitation of the offline estimator for $\eta$ in \Sys is that there is always a possibility that the estimate is off when the system is deployed in the wild. To deal with this scenario, the accuracy of $\eta$ could be assessed at runtime, and then $\eta$ could be updated via an online or an offline re-estimation process. Such an assessment is possible since $\eta$ is used by the system to predict the energy state of the next time slot, for which, the system observes the ground truth immediately after the current time slot, and thus, the system can precisely compute the prediction error at runtime. Depending on the amount of this error, $\eta$ can be adapted to $\eta \pm \delta \eta$, where $\delta \eta$ is proportional to the prediction error.



In our experiments, however, we do not have to perform such adaptations as the empirically derived values for $\eta$ have been fairly accurate. Here, the accuracy of estimation refers to how closely we are able to characterize the randomness in intermittent energy, as opposed to accurately predicting the harvested energy. In Figure~\ref{fig:eta_val}, we show that the estimated value of $\eta$ for three harvesters (used in Section~\ref{sec:energy_study}) converges to their respective prediction accuracy values. For example, the kinetic energy harvester's estimated $\eta$-factor is 0.65, and its (measured) accuracy of predicting the energy state of the next slot is also close to 65\%. The convergence of these two values indicates that the estimate is fairly accurate.

\subsection{Limitations of the \Sys Scheduler}
Although the scheduler in \Sys outperforms state-of-the-art scheduling techniques, it has some limitations that need further investigations. First, the scheduler does not provide any guarantee that all the jobs will finish their mandatory part before the deadline. This is primarily due to the uncertainty of the intermittent energy which does not allow us to formally approach the scheduling problem without introducing any probabilistic terms. Besides, in this paper, we only provide a necessary condition for schedulability analysis (Section~\ref{sec:condition}), while deriving a sufficient condition remains an open problem. Second, the current design of \Sys does not schedule the wake-up cycles of the system. Instead, it reactively wakes up (and shuts down) based on the harvested energy/input voltage. Because of this, the system often misses capturing the events as it might be in power down state. By learning the event pattern and incorporating the probability of job arrivals into the scheduling framework, this problem can be addressed. We leave it as one of our future works.  
Third, the queue size has a significant effect on the scheduler. Due to memory limitations, we cannot implement a longer queue (in Section~\ref{sec:eval}, the queue size is 3). If the queue size is smaller (e.g., 1), the scheduler will only schedules the mandatory portions.


\section{Conclusion}
In this paper, we introduce a deadline-aware DNN runtime framework for intermittent systems. First, We devise a generic metric, the $\eta$ factor, that models the predictability of an energy harvesting system. Second, we propose a DNN construction and execution technique which adapts the DNN inference process at runtime, and decreases the execution time by 5\%-26\%. Finally, we utilize the $\eta$ factor and the adaptive execution framework of a DNN to devise an online scheduling algorithm for batteryless systems that successfully schedules 9\%-34\% more tasks than traditional scheduling algorithms. We also derive a necessary condition for scheduling real-time imprecise DNN tasks on intermittently-powered systems.


\section*{Acknowledgments}
This paper was supported, in part, by NSF grants CNS-1816213 and CNS-1704469.

\bibliographystyle{ACM-Reference-Format}
\bibliography{bibliographies/deep-learning,bibliographies/intermittent-computing,bibliographies/memory-management,bibliographies/realtime,bibliographies/batteryless-detector,bibliographies/network,bibliographies/extra,bibliographies/audio}


\end{document}